\documentclass[review,11.5pt]{elsarticle}
\usepackage{setspace}
\usepackage{ulem}

\usepackage{hyperref}

\usepackage{natbib}
\usepackage{hyperref}
\usepackage[utf8x]{inputenc} 
\usepackage{graphicx}
\usepackage{float}
\usepackage{subfigure}
\usepackage{multirow}
\usepackage{array}
\usepackage{footnote}
\usepackage{color}
\usepackage{ulem}
\usepackage{geometry}
\newcommand{\PreserveBackslash}[1]{\let\temp=\\#1\let\\=\temp}
\newcolumntype{C}[1]{>{\PreserveBackslash\centering}m{#1}}
\newcolumntype{R}[1]{>{\PreserveBackslash\raggedleft}m{#1}}
\newcolumntype{L}[1]{>{\PreserveBackslash\raggedright}m{#1}}

\begin{document}
\DeclareUnicodeCharacter{8208}{ }

\begin{frontmatter}

\title{How do online consumers review negatively?}

\author[mymainaddress]{Menghan Sun}

\author[mymainaddress,mysecondaryaddress]{Jichang Zhao\corref{mycorrespondingauthor}}
\cortext[mycorrespondingauthor]{Corresponding author: jichang@buaa.edu.cn}
\address[mymainaddress]{School of Economics and Management, Beihang University, Beijing, China}
\address[mysecondaryaddress]{Beijing Advanced Innovation Center for Big Data and Brain Computing, Beijing, China}

\begin{abstract} 
Negative reviews on e-commerce platforms, mainly in the form of texts, are posted by online consumers to express complaints about unsatisfactory experiences, providing a proxy of big data for sellers to consider improvements. However, the exact knowledge that lies beyond the negative reviewing still remains unknown. Aimed at a systemic understanding of how online consumers post negative reviews, using 1,450,000 negative reviews from JD.com, the largest B2C platform in China, the behavioral patterns from temporal, perceptional and emotional perspectives are comprehensively explored in the present study. Massive consumers behind these reviews across four sectors in the most recent 10 years are further split into five levels to reveal group discriminations at a fine resolution. Circadian rhythms of negative reviewing after making purchases were found, and the periodic intervals suggest stable habits in online consumption and that consumers tend to negatively review at the same hour of the purchase. Consumers from lower levels express more intensive negative feelings, especially on product pricing and seller attitudes, while those from upper levels demonstrate a stronger momentum of negative emotion. The value of negative reviews from higher-level consumers is thus unexpectedly highlighted because of less emotionalization and less biased narration, while the longer-lasting characteristic of these consumers’ negative responses also stresses the need for more attention from sellers. Our results shed light on implementing distinguished proactive strategies in different buyer groups to help mitigate the negative impact due to negative reviews.
\end{abstract}

\begin{keyword}
online reviews\sep user levels
\sep negative reviewing \sep consumer behavior \sep e-commerce
\end{keyword}
\end{frontmatter}

\section{Introduction}
\paragraph*{}
E-commerce has developed rapidly in recent decades, and it has been estimated that there will be at least 16 billion people purchasing online, representing more than 50\% of netizens around the world in 2018 \cite{ECommerceGlobal}. Among various forms of e-commerce, online platforms are perceived as a concentrated reflection of national consumption power and a societal development trend. Digital traces, generated by massive online consumers, accumulate in these platforms and offer a new probe for the collective behaviors of consumers. Most of the existing studies \cite{Danescu-Niculescu-Mizil:2009:ORO:1526709.1526729,Linden2003,Paolacci2010,Berinsky2012} that focused on the e-commerce area are based on datasets from Amazon in developed countries. In contrast, China, with the largest potential market and highest expansion speed in the world, at 772 million and 14.3\%, respectively \cite{ECommerceGlobal,ECommerceReport}, has been seldom afforded substantial attention in previous exploitations. In China, online buying is experiencing a rapid growth, and over 74.8\% of Internet users now purchase online \cite{InternetReport}, suggesting a high penetration rate of the top e-commerce platforms such as JD.com. In the meantime, considering that the total number of online consumers in China has already exceeded the entire population of the USA, while its increasing rates of netizens and online consumers are positioned in the forefront of the world \cite{ECommerceGlobal}, samples from JD.com would reasonably offer a solid and reasonable base for the behavioral understanding of online reviewing.
\paragraph*{}
Among many of the features of e-commerce, the online review, in a mixed form of a quantitative score and a qualitative text content, is sometimes completed with attached pictures and dialogs with sellers and is always receiving great attention in both academia and business studies, showing its inherent capability to help in consumer relationship management (CRM). In particular, it has been extensively proven that negative reviews, namely, those with poor ratings, are unexpectedly perceived as more helpful \cite{Rozin2001} and persuasive \cite{Chen2013} to other consumers, suggesting their profound impact on future sales \cite{Ghose2011,Berger2010}. Even more important are the close connections among negative reviews and the purchase conversion of consumers \cite{Lee2018}, product awareness\cite{Sahoo2018} and consumer attitude \cite{Lee2008}, which are also reasonably demonstrated, indicating that these negative reviews could be a new inspiring source of consumer behavior understanding.
\paragraph*{}
The previous studies, though suggesting the value of negative reviews, are mainly focused on negative reviews rather than the negative reviewing. The user-centric questions, such as who often posts negative reviews, when and why consumers post them, what emotions are expressed in negative reviews and how these emotions evolve and shift, still remain unknown. However, answering these questions is indeed beneficial to understanding and profiling users in a more precise manner. Moreover, the characteristics of different consumers\cite{Bai2018,Lee2008} provide the possibility to divide online consumers into distinctive groups, offering a finer resolution in probing the various patterns in behaviors of online reviewing. These indeed motivate our study, which aims at obtaining a comprehensive and deep understanding of user behaviors in negative reviewing on e-commerce platforms, by taking different levels of online consumers into consideration from a large volume of negative reviews from JD.com, the largest and the most influential Chinese B2C platform. Stable yet group-dependent patterns beyond negative reviewing will be thoroughly explored in temporal, perceptional and emotional views separately.
\paragraph*{}
In this paper, 1,450,000 negative reviews within 47,290,000 reviews of JD.com are collected. In these reviews, traces of negative reviewing in the most recent 10 years in four sectors are completely collected. Furthermore, a large volume of online consumers, in which 2,050,000 are uniquely identified, are split into five levels. A data-driven solution with various cutting-edge methods is thus employed to mine the dataset and reveal universal behavioral patterns of negative reviewing. We find that the intervals between negative reviewing and online buying are periodic, with cycles of 24 hours or multipliers thereof. The circadian rhythm of negative reviewing essentially suggests the timing consistence in online consumption. It is interesting that consumers of more adjacent levels behave more similarly in negative reviewing. Lower-level consumers are more sensitive to product pricing and the attitudes of sellers, while the upper levels complain more about the quality of the product function and customer service. Moreover, from the perspective of emotion, users of lower levels surprisingly express more intensive negative feelings, but those of the upper levels show a stronger momentum of negativeness. This is the first time, to our best knowledge, that a comprehensive probe of the behavioral patterns in posting negative reviews has been undertaken.
\paragraph*{}
This article provides a comprehensive and systemic understanding of user behaviors in negative reviewing that has previously not received enough attention in both academia and practice. In terms of academics, this paper creatively focuses on the differences across user levels to characterize negative reviewing in a more precise and distinguished way. From the practical perspective, it provides guidelines for precise marketing and profiling and marketing according to user levels to mitigate the negative impact of negative reviews and promote sales, consequently improving service quality.
\paragraph*{}
The remainder of this article is organized as follows. In Section 2, we summarize the related literature to demonstrate the motivations of the present study. The dataset and preliminary methods employed will be introduced in Section 3. The results of our study are described in Section 4, including patterns in temporal, perceptional and emotional aspects. Section 5 and Section 6 provide discussions and concluding remarks for future research.

\section{Related Work}
\paragraph*{}
Driven by multiple factors, such as industry, business, and society, e-commerce has been greatly boosted, creating opportunities and providing various benefits to organizations, individuals and society \cite{Turban2008}, even guiding the formation of a new lifestyle and economic situation in recent years. Therefore, plenty of studies have focused on e-commerce due to its vital roles in economics. 
\paragraph*{}
Extensive studies have contributed to the advantages \cite{Quayle2002}, challenges \cite{Quayle2002} and patterns among different cultures \cite{Gefen2006}, aiming to determine the development model of e-commerce or enhance techniques to help corporations in its operation and CRM, such as data mining methods \cite{Ahn2010,Lee2010,Chen2009} in understanding consumer purchase behavior. He et al. employed text mining approaches to model the business competition environment of a pizza seller with social media comments \cite{He2013}, and Huang et al. examined the impact of social media interaction on consumers' perceptions in online purchases \cite{Huang2017}. In addition, interacting factors and techniques implemented in e-commerce platforms have been studied from various perspectives and dimensions. Much attention has been paid to the inner reactions and features of e-commerce, for instance, purchase return prediction \cite{Sahoo2018}, customer engagement \cite{Yang2014}, helpfulness of reviews \cite{Mudambi2010}, various identities of online consumers \cite{Koufaris2002}, characteristics of different consumers \cite{Lee2008} and advertisement features \cite{Lee2018} to promote consumer conversion. However, there is still much room to continue digging into inner reactions and human behaviors in e-commerce. Meanwhile, there is no shortage of empirical exploitations based on specific e-commerce platforms, such as Amazon \cite{Danescu-Niculescu-Mizil:2009:ORO:1526709.1526729,Linden2003,Paolacci2010,Berinsky2012}, JD.com \cite{Wang2010}, Taobao \cite{Ye2009,Li2008} and eBay \cite{Li2008}. China, with 25\% of the global population and the largest market in the world, has been officially reported to have gained an 11.7\% year-on-year growth in national e-commerce transaction volume and 14.3\% in the national online shopping user scale in 2017 \cite{ECommerceReport}. Considering the surge of e-commerce in China, along with the vast amount of online consumers in China and its highest rate of increase in netizens and online consumers worldwide \cite{ECommerceGlobal,InternetReport}, this article aims to provide a reasonable yet profound insight into consumers' behavior in e-commerce, based on data from Chinese e-commerce. Furthermore, compared with previous studies that aimed at discerning how e-commerce influences consumers, the perspective of potential feedback that e-commerce can acquire from consumers' behavioral patterns is lacking and needs attention.
\paragraph*{}
Among various topics of e-commerce research, serving as kind of word-of-mouth (WOM) concept, negative reviews are always a topic of great concern, as users always express their complaints and negative feelings about unsatisfactory experiences through online reviews after a purchase. In the trend of user-generated content, the negative review is an ideal channel for online sellers to mine consumers' perspectives, then help the e-commerce platform to discover problems and accordingly improve service. Surveys have pointed out that most online consumers browse online product reviews before making shopping decisions \cite{Mcauley2016}, and their attitudes toward a product can be influenced by related reviews or WOM \cite{Wooten1998}. The different features of negative reviews or negative WOM have been primarily explored, suggesting the unexpected value of negative reviews in the business model of e-commerce. Existing efforts have focused on the reference of reviews and proven that negative reviews are more helpful \cite{Rozin2001,Chen2013} for consumers and are regarded as more persuasive \cite{Herr1991} to help purchase decisions. Different from positive WOM, negative WOM demonstrates a stronger nature in the expression of emotion as an intent \cite{Sweeney2012} and is more emotional \cite{Verhagen2013}, which can serve as a proxy to monitor consumers' feelings. It has been disclosed that review extremity \cite{Mudambi2010} and questioning attitude \cite{Huang2017} in reviews can affect their helpfulness. In addition, Richins and Marsha suggested that if a negative review is not addressed in a timely manner by the seller, its negative effect will spread to a wider range and cause harm to the company \cite{Richins1983}. Nevertheless, positive outcomes of negative reviews are also surprisingly unraveled. Ghose et al. investigated the relationship between the objectivity of a review and product sales or helpful votes and pointed out that negative reviews with more objectivity obtain more votes \cite{Ghose2011}. Berger et al. further provided evidence that negative reviews may even increase sales by increasing awareness \cite{Berger2010}. Therefore, the studies above highlight the value of negative reviews and imply the need for a systematic understanding of behavioral patterns beyond negative reviewing from the perspective of online consumers; however, less attention has been paid to this perspective in existing exploitations. 
\paragraph*{}
In fact, modeling online user behavior can help e-platforms target customers and implement market segmentation \cite{Dibb1998} and marketing strategy \cite{Liao2009}. Koufaris applied a technology acceptance model and flow theory to model online consumers and identified the online consumer's dual identity with a buyer and a computer user \cite{Koufaris2002}. Tan et al. introduced sentiment analysis and network structure analysis into user-level feature identification \cite{Tan:2011:USA:2020408.2020614} but ignored the preferences of consumers. Intuitively, while Ghose et al. aimed at the user's status in the e-commerce network and modeled the helpfulness of reviews with the historical data of the user \cite{Ghose2011}, the study did so without the delineation of the overall or stable features of individuals. Bai et al. focused on characteristics of early reviewers and their benefits for marketing \cite{Bai2018}. Lee et al. divided online users into high-engagement and low-engagement groups and concluded that the two groups have different performances in emotional perception and purchase desire \cite{Lee2008}, providing the possibility to further divide groups based on consumers' characteristics. However, while digging into negative reviews, the existing studies on buyer behavior inherently miss the focus of negative reviewing, such that the various patterns that depict how consumers of different groups negatively review still remains unknown.
\paragraph*{}
A review of the relevant research on negative reviews and the behavior among the sources that we accessed revealed that the insightful value of negative reviews has been illustrated; however, we were unable to find research focused on understanding the buyer behavior that underlies negative reviews. Moreover, in related research, a fine resolution is not reached when distinguishing different users, and even worse, the considered features lack richness. Motivated by these issues, this paper aims to shed light on online consumers' behavior patterns of negative reviews from the perspective of various dimensions and a fine resolution of user levels.
\section{Materials and methods}
\subsection{Dataset overview}
\paragraph*{}
In this study, we focus on e-commerce platforms in China, which has the largest e-commerce marketing on the scale of consumers and related firms and has the highest rate of increase in online consumers, to probe into various behaviors beyond negative reviewing. Therefore, we implemented our research based on an online review dataset from JD.com, which is the largest B2C platform in China and increased in market size and gross merchandise volume at the rate of 0.8\% and 30.4\%, respectively, in the first quarter of 2018.
\paragraph*{}
An overview of the dataset we used in this study can be found in Table \ref{Table 1}, and a brief description of the attributes is shown in Table.A1. The dataset will be publicly available through a permanent link of Figshare repository later. In total, we collected 1.45 million negative reviews within 47 million reviews of four sectors, and the time span ranges from 2008 to 2018, covering the most recent 10 years. Due to anonymity, only 2.05 million users that are associated with these reviews can be uniquely identified. In addition, the rich attributes of online reviews on JD.com guarantee the feasibility of measuring various behaviors of reviewing in our following experiments. The four sectors selected out of all 14 sectors are the main commodity categories on JD.com and can cover different commodity characteristics to obtain universal patterns. It should be mentioned that there are few spam reviews or bot reviews in our dataset due to the comment mechanism on JD.com, in which sophisticated techniques such as captchas are extensively employed to prevent false reviews. In the meantime, being a self-operated platform \cite{wu2019online}, there is no motivation for JD.com to encourage paid reviews. Moreover, we deduplicated the dataset according to the comment ID, a unique identifier of reviews.
{}
\begin{table}[h]
\footnotesize
\caption{An overview of the dataset used in this study.}
\centering
\begin{tabular}{m{2cm}|R{1cm}|R{1.5cm}|R{1.5cm}|C{3cm}|R{1.3cm}}
\hline
Sector & \# Subsectors  & \# Reviews  & \# Negative Reviews  & Time Span & \# Identified Users \\
\hline
Computers  & 16                    & 15,130,000          & 480,000                     & 2008.11.02-2018.03.22 & 620,000                       \\
\hline
Phone \& Accessories & 12                    & 17,150,000          & 500,000                     & 2008.11.21-2018.03.06 & 950,000                       \\
\hline
Gifts\&Flowers   & 10                    & 2,350,000           & 100,000                     & 2011.02.13-2018.04.07 & 70,000                        \\
\hline
Clothing          & 23                    & 12,660,000          & 370,000                     & 2011.05.07-2018.04.07 & 410,000                       \\
\hline
\end{tabular}
\label{Table 1}
\end{table}

\paragraph*{}
The definition of negative reviews in e-commerce platforms may differ and mainly refers to reviews expressing consumers' negative evaluations, usually evaluated as low scores. In terms of the dataset from JD.com in our study, negative reviews refer to the one-score (the lowest score) reviews from consumers. The evaluation system for consumer reviews on JD.com is similar to that on Amazon.com---consumers who bought certain goods can post their comments in text and photos and a score about their purchase experience. Thus, reviews are regarded as a generalized index of reviewers' attitudes and served as an ideal proxy to probe buyer behaviors online in this paper.

\paragraph*{Overview of Negative Reviews}
An overview of the negative review dataset can be seen in Fig. A1. Fig. A1(a) shows the proportion of different scores in our dataset, and it can be seen that the negative reviews only occupy a small percentage ($<$0.05). Even though the number of negative reviews is small, to some extent, their significance is greater than that of positive reviews with higher scores since operation or management problems are exposed in negative reviews and often receive more helpful votes, such that the attribute usefulCount can serve as an indicator \cite{Rozin2001,Chen2013} (see Fig. A1(b)).
\paragraph*{Overview of Users}
Fig. A1(c) presents an overview of users and user levels in our dataset from the aspects of anonymous and identified proportions, and Fig. A1(d) shows the percentage of different scores in different user groups, distinguishing the four sectors. Though the identified consumers occupy a relatively smaller amount than the anonymous ones, the total volume of the four sectors still approaches 2.05 million, far exceeding the population size in conventional surveys or questionnaires. Thus, it will reliably facilitate the subsequent trace of emotion evolution of negative reviewing at the individual level. However, for the examination at the collective level, the individual identification of consumers will no longer be necessary. The attribute of the user level of the review (see Table A1) can be directly used to divide consumers into different levels. Specifically, from copper and silver to golden and diamond, as the user level increases, it indicates that the user has purchased more or that the purchase expense is higher, and PLUS users can be seen as VIP users who paid an extra fee to enjoy more membership benefits.

\subsection{Method}
\paragraph*{}
To characterize online consumers' behavior of negative reviewing from multiple aspects, the texts of negative reviews offer rich signals, and several cutting-edge text mining methods will be employed here. Specifically, topic classification, aspect mining, word embedding and semantic distance measuring will be used to probe various reviewing behaviors.
\paragraph*{Preprocessing}
In each sector, we identify duplicate reviews by their unique identifiers and remove them from the dataset. The segmentation of sentences and filtering of stopping words are executed in sequence, and a list of Chinese terms is formed and ready for feature vector generation and other follow-up processes.
\paragraph*{Classifier Construction}
To recognize what consumers are complaining about in a negative review, we implemented a supervised classifier for automatic reason detection in negative reviews. It is found that product complaints and poor service are always sources of online consumers' unsatisfactory shopping experience and trigger the posting of negative reviews \cite{Craig62,doi:10.1080/13527266.2013.797778}. According to this, after browsing hundreds of negative reviews in our dataset, we consider product complaints to be product defeats, and poor service is further subdivided into poor performance in logistics service, consumer service and marketing service. Therefore, we determined the four factors that lead to negative reviews online, as outlined below:
\subparagraph*{}
\textbf{Logistics}: reviews complaining about the unsatisfactory performance related to logistics, such as slow logistics, slow delivery, bad attitude of courier service staff, parcel damage during delivery, etc.; \textbf{Product}: reviews expressing defeats about the product itself, which can differ from categories of the product, for example, negative reviews about computers may contain comments about poor CPU performance and heat dissipation; \textbf{Consumer service}: reviews whose content is about customer service, such as lack of feedback response, problems with installation service, slow solutions to problems, rude attitudes of sellers, etc.; and \textbf{Marketing}: reviews about the sellers' false marketing behavior, such as untrustworthy or malicious price increases or false advertisements.
\subparagraph*{}
Regarding the details of the classifier's construction, 8153 negative reviews randomly selected from four sectors in our dataset, were treated as training data, and tagged by five well-trained coders separately, who have rich experience in both online shopping and reviewing. Various algorithms were applied to train the classifier, where logistic regression algorithm and the bag-of-words model produced the best performance, with a word list as input and four reason labels as output, detailed as Logistics, Product, Customer service and Marketing. Specifically, the five-fold cross-validation of the classifier demonstrated that its accuracy and recall both exceeded 0.80, which is sufficient for the following analysis.
\paragraph*{Aspect mining}
To further divide reasons into more concrete topics and obtain a more detailed understanding of why users post negative reviews, we applied a latent Dirichlet allocation (LDA) \cite{Pritchard945,Blei2003} model to mine aspects. As a generative statistical model, it allows for sets of observations to be explained by unobserved groups, which helps automatic topic recognition and is now widely applied in natural language processing. Moreover, the advantage of automatic clustering and manpower savings make the LDA algorithm suitable for situations where concrete aspects are unknown. With respect to the parameter settings, the topic number, topic similarity and perplexity will be comprehensively considered.
\paragraph*{Word Embedding}
To better reflect the semantic similarity in the background of online shopping, we have to first train a word-embedding model on the negative reviews collected. Word2vec is a group of models where words or phrases from vocabulary are mapped to vectors of real numbers \cite{NIPS2014_5477}. In this study, considering that the length of texts in negative reviews is relatively short, the continuous bag-of-words (CBOW) algorithm, which can effectively enrich the semantic density and overcome the possible sparsity, is employed to train the embedding model, with a minimum count of one and a vector size of 200. The output of word vectors is used in the word net construction and the semantic distance calculation that follows.
\paragraph*{Semantic distance}
Consumers may have different expression habits in negative reviewing, especially considering the various user levels. It is assumed that habits of expression in reviewing could be essentially reflected by the co-occurrence of words in texts, i.e., the similarity of word pairs, in which more similar words would co-occur more frequently. According to this assumption, the drifting of expression habits across levels can be well measured through the variation of similarities between common words that are adopted by all levels. To measure the similarity landscape, a word network of semantic closeness will be first established for each user level. Specifically, for negative reviews posted by users of a certain level, a word2vec model is trained, and the semantic similarity between words can then be calculated by the cosine distance between the corresponding embedding vectors that are derived from the embedding model. In building the word network of a certain level, each word in common words will be connected to its most similar $N$ neighbors inferred from the word2vec model of the corresponding level, where $N$ can be tuned to adjust the threshold of semantic closeness. Then, for the drifting of expression habits of two levels, we simply rank common words according to their structural indicators such as degree, k-core index and clustering coefficient in level-dependent word networks and employ the Kendall rank correlation to represent the variation of the semantic landscape across these two levels. For example, larger correlations indicate more consistent expression habits.
\paragraph*{Emotion Measures} 
To measure the emotions delivered in reviews, a dictionary-based solution \cite{DiWeng} is selected, which contains 1180 positive Chinese words and 1589 negative Chinese words. Two emotion indexes are further defined. The positive (negative) rate denoted as $r_{pos}(r_{neg})={n_{pos}(n_{neg})\over n}$,
can be calculated by the proportion of positive words or negative words in this review, where $n_{pos}$ or $n_{neg}$ refers to the number of positive or negative words in a review text, and $n$ refers to the number of total words in a review text. Furthermore, the polarization, defined as
$$i_{polar}=\left\{
\begin{array}{rcl}
{n_{pos}-n_{neg}}\over{n_{pos}+n_{neg}} & , & {n_{pos}+n_{neg} \neq 0} \\
0 & , & {n_{pos}+n_{neg} = 0}
\end{array} \right.$$
indicates the extent of polarization, for example, positive implies positive emotion, negative implies negative emotion, 0 is neutral, and approaching -1 of the polarization indicates that the review is extremely negative.

\paragraph*{}
It is worth mentioning that all of the following analyses will be conducted from two views, in which the collective view means putting all users together, and the user-level view means grouping users into levels. The level of a certain user can be recognized according to the attribute of 'UserLevel' in our dataset.

\section{Results}
\paragraph*{}
In this article, to understand online consumers' negative reviewing behavior from a comprehensive perspective, consideration is given to the following: when consumers would like to post negative reviews, such as the exact time in a day or the interval after they bought a product, which is concluded as temporal dimension; what content they express in negative reviews or why they post negative reviews, which can be summarized as a perceptional dimension; and how they express the negativeness and how their emotion evolves, which are generalized to the emotional dimension.
\subsection{Temporal Patterns}
\paragraph*{}
The temporal patterns of negative reviewing will be probed from the review-creation (RC) time and intervals between the review-creation time and the product-purchase (PP) time. 
\subsubsection*{Review-creation time}
\begin{figure}[h!]
\includegraphics[height=7.0cm]{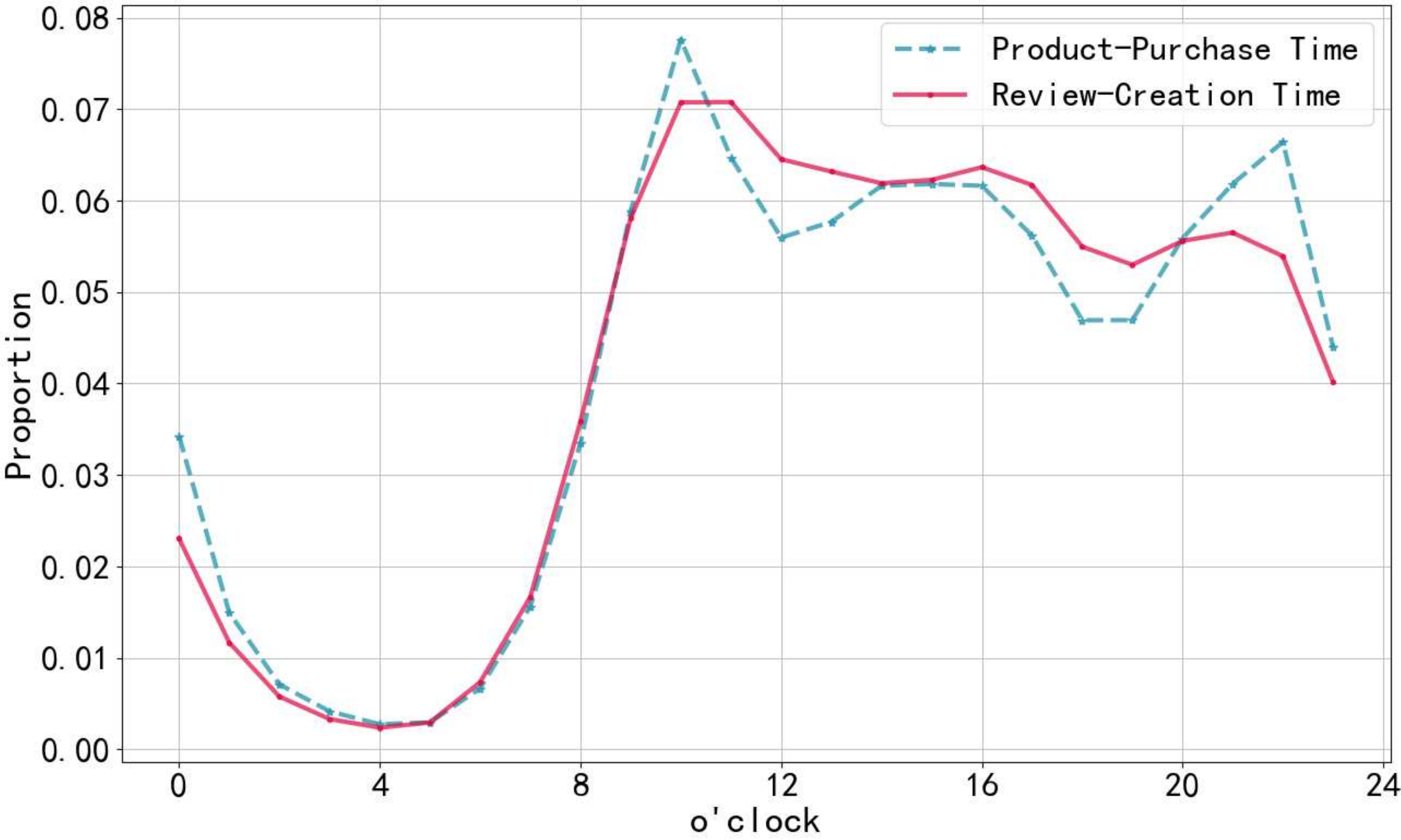}
\centering
\caption{\textbf{Distribution of review-creation time and product-purchase time on an hourly scale.All data represent negative reviews.} All data represent negative reviews. The x-axis represents the hour, and the y-axis represents the proportion of consumers that purchase or post negative reviews at the corresponding hour.}
\label{Fig. 2}
\end{figure}

\paragraph*{}
The distribution of review-creation time on an hourly scale can be seen in Fig. \ref{Fig. 2}, which contrasts with that of the product-purchase time. As shown in Fig. \ref{Fig. 2}, the posting pattern of negative reviews in the temporal aspect is consistent with most online behaviors, such as the user active time \cite{PhysRevE.78.026123} or the log-in time on social media \cite{Guo2011}. In addition, we could not identify obvious differences between the distributions of the review-creation time and product-purchase time, unless there is a minor lag in RC time compared to PP time. It seems that there is more shopping in the morning but more negative reviewing at night. Moreover, this lag also inspired further examination of the interval between RC time and PP time.

\subsubsection*{Intervals between RC time and PP time}
\begin{figure}[h]
\centering
\includegraphics[width=12cm]{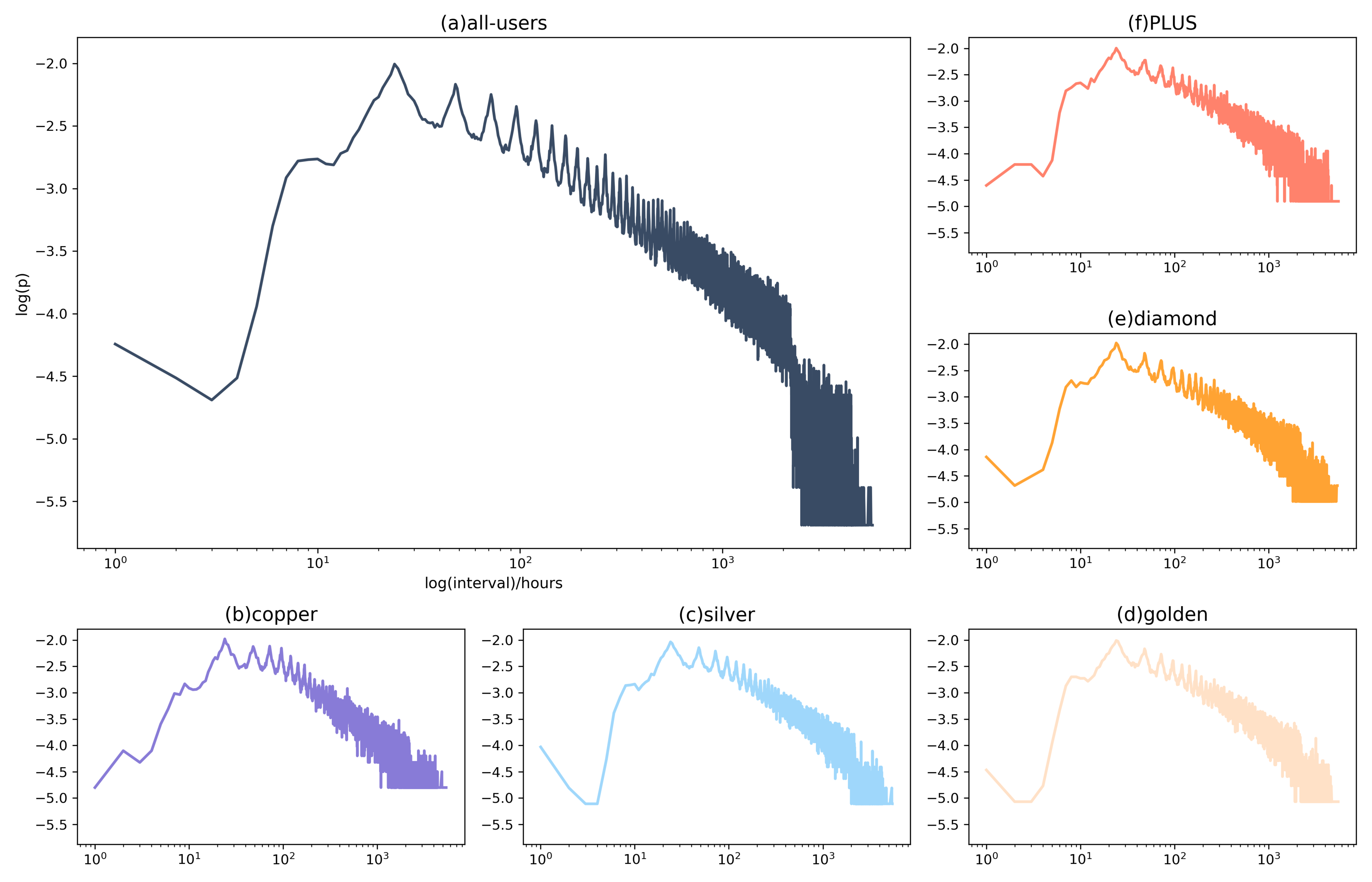}
\caption{\textbf{Log-log distributions of intervals (minutes) for RC and PP time, taking the sector Computers as an example.} The subplot (a) is for the all-user group. Subplots from (b)-(f) correspond to a certain user level, from copper to PLUS. For every subplot, the x-axis is the interval on a log scale, and the y-axis is the proportion on a log scale.}
\label{Fig. 3}
\end{figure}

\paragraph*{}
There is always a time difference, i.e., an interval, between RC time and PP time, and it suggests a process of product or service experience and usage of consumers. The exploration into intervals between RC and PP time helps understand consumers' rhythm in negative reviewing from both general and user-level perspectives.
\paragraph*{}
For every purchase record from different sectors in our dataset, we obtained the interval in hours and its log-log distribution, as shown in Fig. \ref{Fig. 3}, which contains distributions for all five user levels. As presented, the overall form corresponds to heavy-tail-like and power-law-like distributions, which is consistent with most online human dynamics \cite{Wang2010,PhysRevE.78.026123,Guo2011}, indicating that negative reviewing after the purchase is bursty \cite{Barabasi2005}. In addition, comparing the all-user group with different user-level groups, no significant difference can be found, and the five user levels all perform according to a heavy-tail distribution.

\begin{figure}[h]
\centering
\includegraphics[width=12cm]{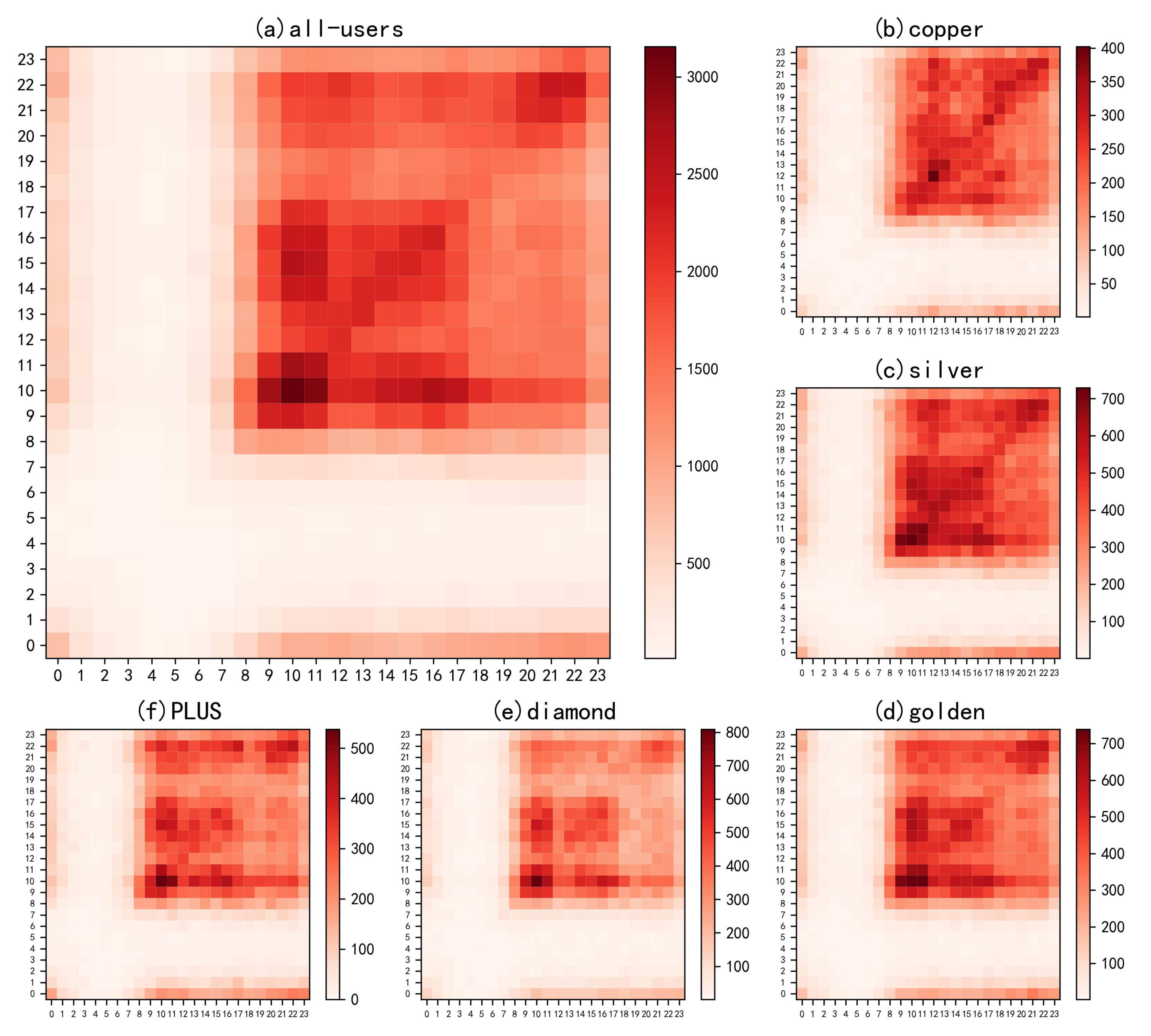}
\caption{\textbf{Heatmaps of hours for RC and PP of negative reviews, taking the sector Computers as an example.} The x-axis refers to the hour of PP, and the y-axis refers to that of RC. The hour ranges from 0 to 23. Every patch has an RC and PP pair that it corresponds with, and the color in it refers to its frequency, i.e., the darker red means a greater frequency. The results of other sectors are consistent.}
    \label{Fig. 4}
      \end{figure}

\paragraph*{}
At the same time, in each distribution of Fig. \ref{Fig. 3}, a periodic fluctuation is observed when the interval is larger than 10 minutes. Though the exact distance in the distribution becomes narrow because of the log-scale, a rough measurement surprisingly shows that the fluctuation cycle is approximately 24 hours, with the first peak at 24, which means the interval between RC and PP time is more likely to be hours in multiples of 24, i.e., one day. This interesting phenomenon suggests a circadian rhythm in negative reviewing.
\paragraph*{}
To further examine the reliability of the periodic intervals, we thoroughly check the exact correlation between the purchase hours and the reviewing hours of all negative reviews. As illustrated in Fig. \ref{Fig. 4}(a), for all users, the occurrence of pairs with the same or close PP and RC hour, located at or near the diagonal, is significantly higher than that of other cases. It is also interesting that as pairs of the RC and PP hour approaches the diagonal, i.e., reviewing at the same time of buying, the occurrences demonstrate a significant increase. This finding indicates the reliability of the periodic intervals and its cycle of 24 hours, which can be interpreted as online consumers tending to post negative reviews for a certain product at the same hour of buying. 
\paragraph*{}
Comparing the performances among user levels, in Figs. \ref{Fig. 4}(b)-(f), it can be further observed that the same patterns of periodic intervals for silver and golden levels are closer to that for all users. However, patches in grids for diamond users and PLUS users are relatively disarrayed, lacking a regular pattern of gradual change. Therefore, it can be summarized that the online interactive behavior, such as purchasing and posting reviews, of users with higher levels can be triggered in a more random manner, especially compared with that of users at lower levels. This can be well explained by the more frequent purchases and more active interactions of higher-level users.

\begin{figure}[h]
\centering
\includegraphics[width=12cm]{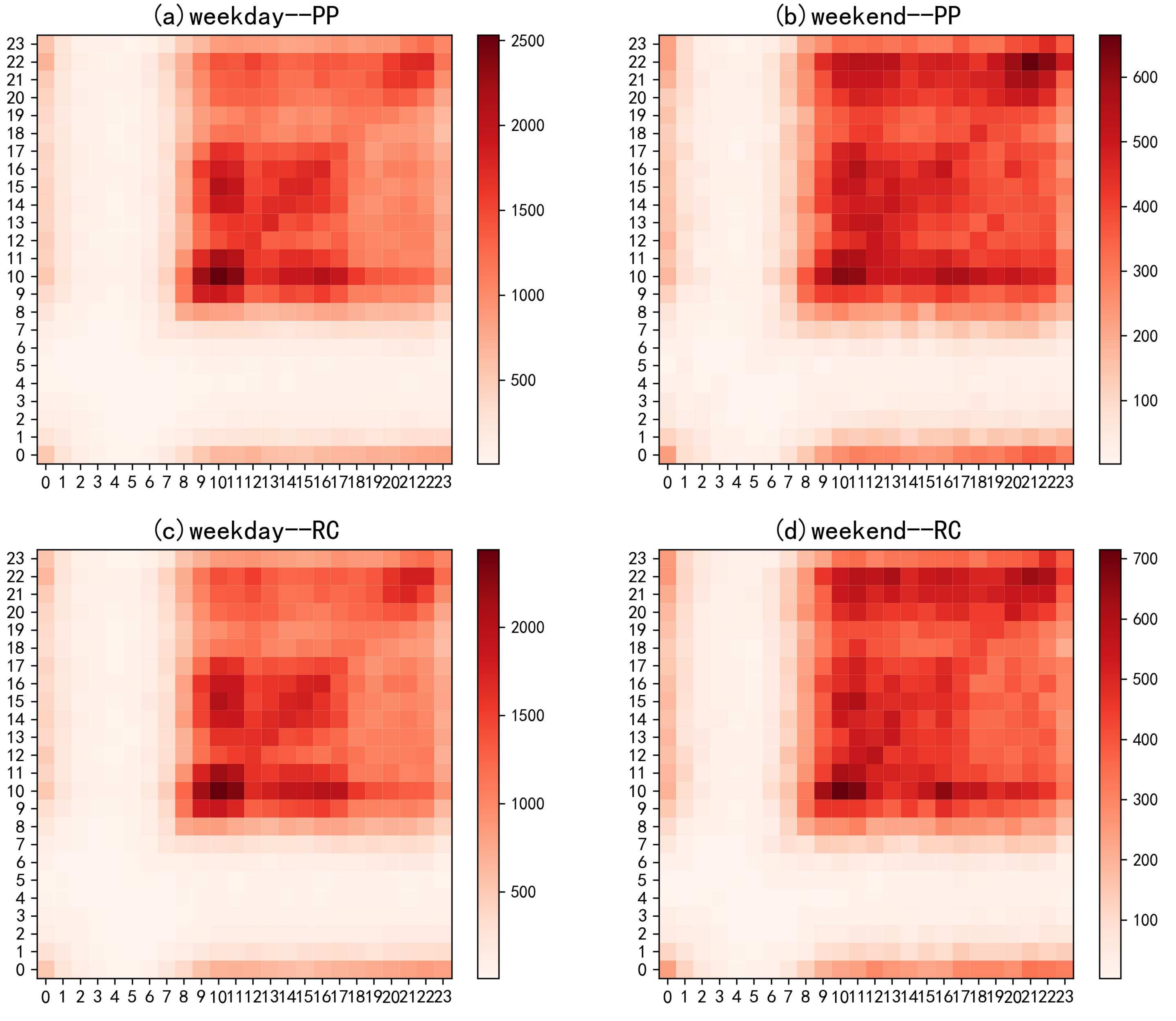}
\caption{\textbf{Different performances in periodic intervals between weekdays and the weekend, taking the sector Computers as an example.} The meaning of x-axis and y-axis are both corresponding with that in Fig. 3. The identification standard of a weekday or weekend is according to PP time in subplots (a) and (b), while it is according to RC time in (c) and (d). The results of other sectors are the same.}
\label{Fig. 5}
\end{figure}

\paragraph*{}
Periodic intervals indicate an interesting relationship between negative reviewing behavior and the purchase, where the purchase is motivated by the user demand, and reviewing comes from user experience after using the product. According to this, we suppose that periodic intervals with a 24-hour cycle might be related to a perceptional rhythm of online consumption, which means that users tend to perform or carry out related activities at a fixed period in a day. To verify this, we implemented an experiment inspired by Yang et al. \cite{Yang:2018:ULP:3146384.3182165}, where the discrepancy of behavior performance on weekdays and the weekend is examined, which implies that randomness and recreation in the behavior during the weekend is much more intense than that of weekdays when people are doing regular work or studying. Fig. \ref{Fig. 5} displays the result of the discrepancy between weekdays and the weekend in periodic intervals and supports our conjecture to some extent. As we can see, the periodic interval of the weekday is more consistent and intensive than that of the weekend, indicating that regular work or study activities enforce the perceptional rhythm, while recreation activities weaken the perceptional rhythm. Therefore, we propose that the regular activities of humans are related to periodic intervals between RC and PP time. 

\begin{figure}[h]
\centering
\includegraphics[width=12cm]{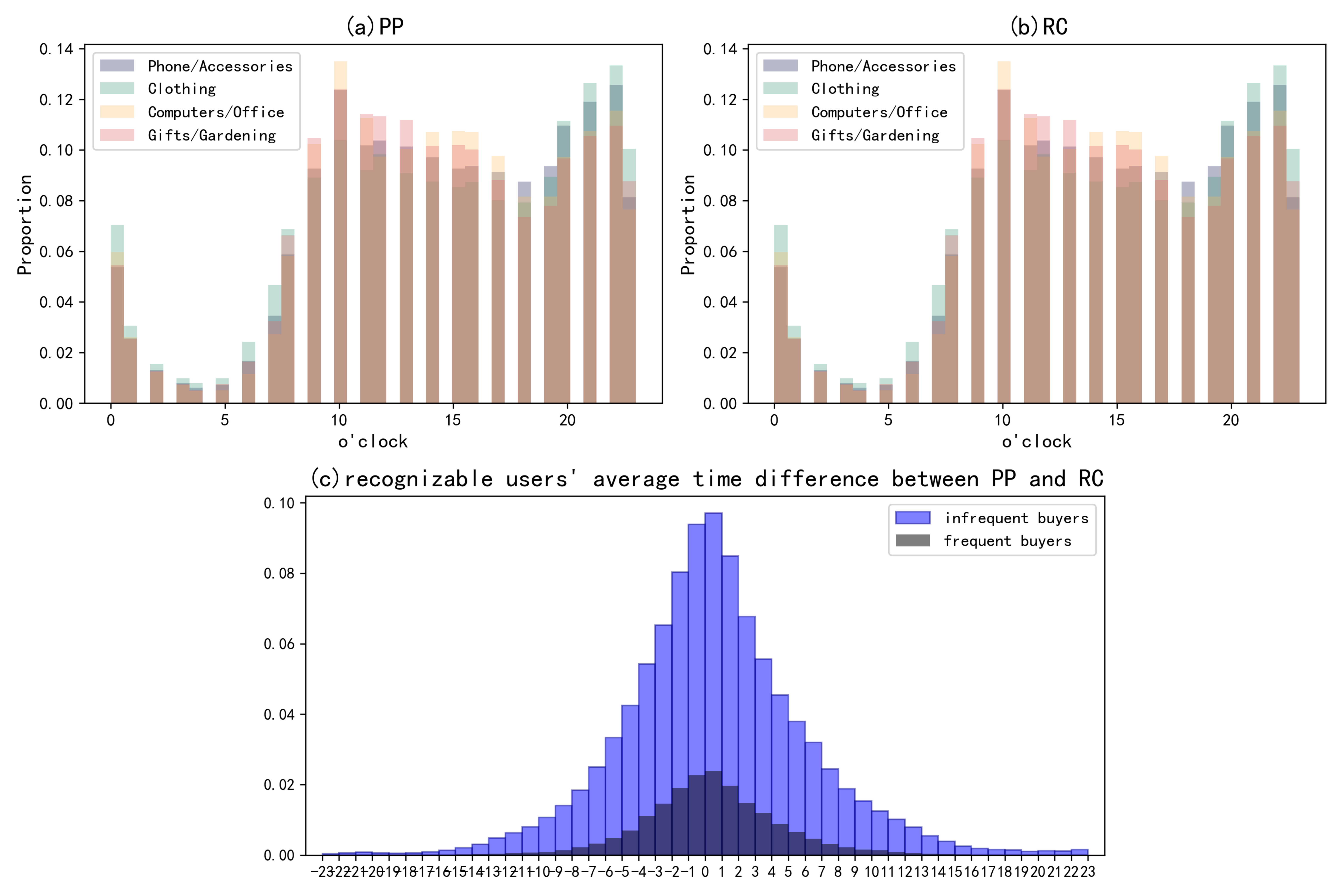}
\caption{\textbf{Distributions of PP or RC time and their intervals derived from uniquely identified consumers on a 2-hour scale.} (a) PP distributions for four different product categories on an hourly scale. (b) RC distributions for four different product categories on an hourly scale. (c) Distribution for identified consumers' average intervals between PP and RC time for all four product categories on an hourly scale. \textit{Here, we filter out identified consumers in every sector who have no less than three purchases in our sampling.}}
\label{Fig. 6}
\end{figure}

\paragraph*{}
Furthermore, as Figs. \ref{Fig. 6}(a)-(b) display, different distributions of PP and RC time in different product categories provide more evidence that perceptional rhythms vary across sectors, which indicates that the time of consumers' purchases and reviewing might be associated with the features of a certain product that might lead to a regular timeline for consumers' online behavior and then to periodic intervals. To provide more detail, we can see in Figs. \ref{Fig. 6}(a)-(b) that the PP time or RC time of products from \textit{Phone\&Accessories} and \textit{Clothing} shows peaks in the evening and early morning, which is always for recreation time. In contrast, \textit{Computers}-related PP and RC times are performed more actively than others at 10 to 16 o'clock, which are normally office hours. 
\paragraph*{}
To enhance the reliability of verification, we implement another experiment from the view of identified users, preventing the statistical bias coming from users who purchase products frequently. In Fig. \ref{Fig. 6}(c), identified users with no less than three purchases are kept to obtain the average hour difference between PP and RC time. Moreover, the proportion of users with more than five purchases (generalized as frequent-purchase users) in all valid identified users is relatively small, which suggests that the high frequency of purchases from frequent-purchase users is not the main factor that leads to periodic intervals. Note that frequent-purchase users accounts for 19.16\% of all users, but generate more than half of all purchases. From Fig. \ref{Fig. 6}(c), we can conclude that periodic intervals come from the stable habits of online consumers in consumption. Therefore, the effectiveness and reliability of circadian rhythms in negative reviewing in Figs. \ref{Fig. 4} and \ref{Fig. 5} and the robustness of periodic intervals can be accordingly testified. Explorations about periodic intervals suggest that the purchase time and reviewing time are sector-dependent and have an interesting relationship with one another, which was rarely discovered previously, and can add elements to marketing strategies, such as combining a timing preference with a user and the product category.
\paragraph*{}
The exploration surrounding the buyer's temporal online behavior and periodic intervals between RC and PP time is suitable for our dataset and aims at its existence and possible causes in this article. The conclusions and conjectures put forward here can provide value for more research into human behavior with perceptional regulation and the possibility for e-commerce practitioners to precisely profile consumers' online active hours or upgrade marketing strategies. 

\subsection{Perceptional Patterns}
\paragraph*{}
One profound value of negative reviews is that they offer a channel to express consumers' dissatisfaction about a product or a service, providing future consumers with reference information and sellers with an direction for improvement. Therefore, recognition of what online consumers complain about in reviews and patterns of how they perceive online buying turn out to be significant. Perceptional patterns refer to the exact reasons beyond consumers' postings of negative reviews and level-dependent preferences related to the inner driving force of cognition toward negativeness. Moreover, it is worth mentioning that differences among consumers' negative reasons might originate from not only the users themselves but also from platform policies for different levels.
\subsubsection*{Main reasons for negative reviews}
\begin{figure}[h]
\centering
\includegraphics[height=7.0cm]{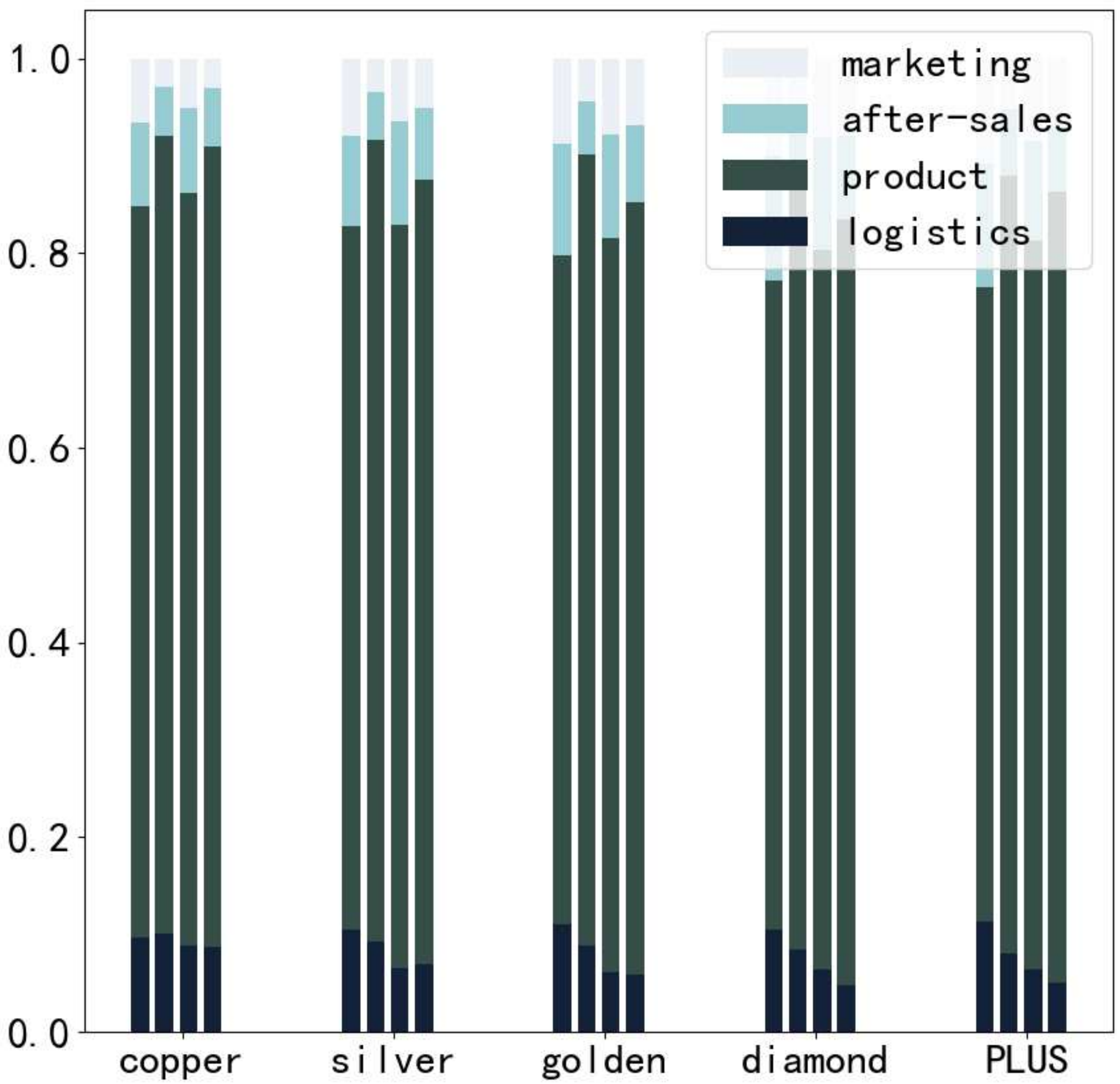}
\caption{\textbf{Proportions of main reasons for negative reviews from different user levels. This is a stacked histogram for the proportion of the main reasons for negative reviews.} The x-axis corresponds to five user levels, and the y-axis corresponds to proportions. Additionally, every four bars with a tiny separation above every user level represent different categories: from left to right, Gifts\&Flowers, Clothing, Computers, and Phone\&Accessories.}
\label{Fig. 7}
\end{figure}

\paragraph*{}
For a medium-sized horizontal e-commerce platform, it can receive several thousand negative reviews that complain about problems from different aspects, of which different departments are in charge. Under this circumstance, it is necessary to automatically identify reasons for a large number of negative reviews, preparing for settlements of the exact problem. Therefore, as discussed in Section 4, we constructed a logistic regression classifier for automatic reason identification of negative reviews.
\paragraph*{}
We apply the logistic regression classifier to all negative reviews in our dataset, and Fig. \ref{Fig. 7} shows the proportion of the four main negative review reasons of five user levels. As can be seen, the sector Gifts\&Flowers receives the largest proportion of complaints about logistics, which is consistent with our common sense that products from this category have high demands for distribution punctuality and professional equipment. Regarding negativeness toward product quality, Clothing ranks first, followed by Computers, and Gifts\&Flowers as the last. The former two have an obvious character of `commodity first'. With respect to dissatisfaction about customer service and false marketing, the ranking order is exactly the opposite, suggesting that the triggers underlying negative reviewing are sector-dependent. Moreover, there is no evident difference among the five user levels, indicating that all consumers perceive the main reasons similarly, and more detailed reasons will be probed in the later analysis.

\begin{figure}[h]
\centering
\includegraphics[width=12cm]{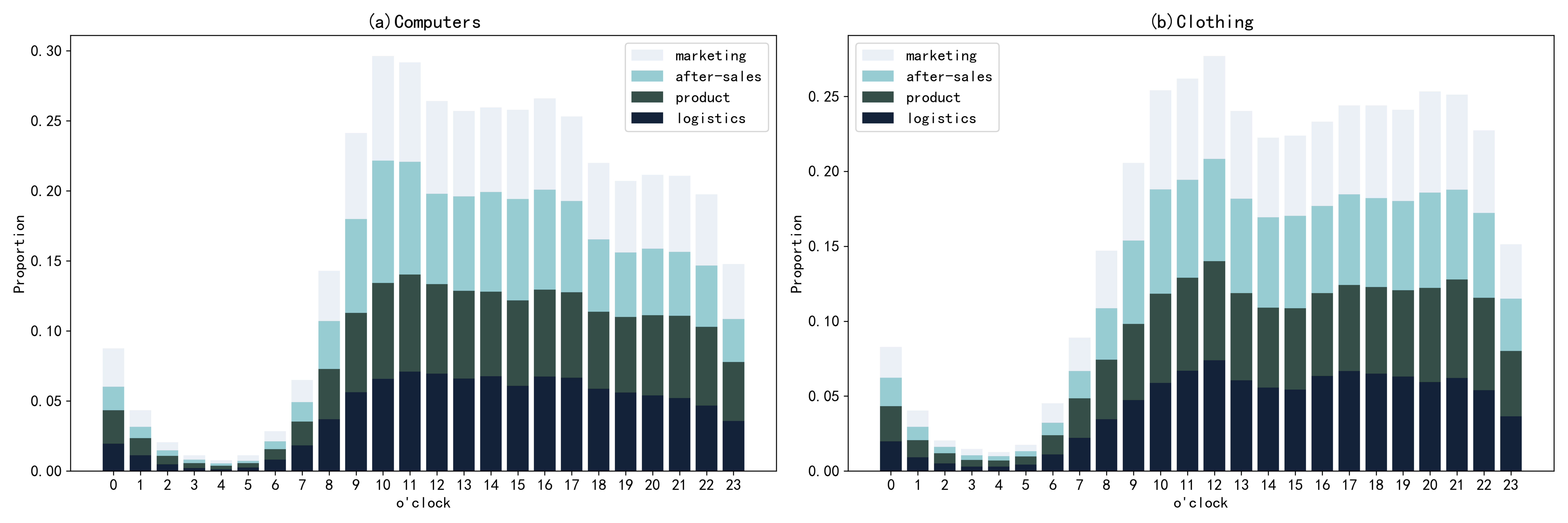}
\caption{\textbf{Proportions of main reasons for negative reviews on an hourly scale.} (a) is the stacked histogram for the sector Computers and (b) is for Clothing. The x-axis is hours, from 0 to 23, and the y-axis is the proportion of a certain reason at the corresponding hour.}
\label{Fig. 8}
\end{figure}

\paragraph*{}
To explore whether there are regular patterns for when to post certain kinds of negative reviews in a day, we follow thoughts from the temporal aspect to focus on time-triggered patterns of reasons for negative reviews. Fig. \ref{Fig. 8} presents proportions of main reasons in an hourly scale for two sectors, which demonstrates different trends. Specifically, the first peak is generally located at 10-12 o'clock as expected. However, the reach of the second peak is sector-dependent, for instance, the sector Clothing, with obvious personal-usage features, reaches its second peak at 20-21 o'clock of the personal time for recreation (the result is similar for the Phone\&Accessories sector, etc., which can be found in Fig. A2), while Computers, with obvious office attributes, misses the second peak. This indicates that even in the hourly pattern of negative reviewing, the reasons that lead to poor reviews are still sector-related.
\paragraph*{}
Comparing the four sectors we selected from our dataset, there are common laws in the time distributions of the main reasons for negative reviews that can be summarized. First, as observed in Fig. \ref{Fig. 8}(b), the negative reviews of logistics obviously reach peaks at 11 to 12 A.M. and 16 to 18 P.M., which is at the pick-up time. Second, there is an upward trend in the number of negative reviews of product quality in the evening, especially for categories with personal use features, such as Phone\&Accessories and Clothing, and this trend is not very evident for sectors with office attributes. Third, a downward trend can be seen in negative reviews that complain about dissatisfaction with customer service for categories with office attributes, which vanishes in categories with personal use features. Moreover, there is no trend observed for negative reviewing due to false advertising. These findings imply that the hourly patterns of posting negative reviews for a certain reason have a close relationship with the schedule of daily life and demonstrate stable rhythms. Moreover, it was not expected that the distribution of the reasons that lead to negative reviewing would be user-level-independent, and no evident discrimination across levels is found.

\subsection*{Detailed aspects leading to negative reviewing}
\begin{figure}[h]
\centering
\includegraphics[width=12.0cm]{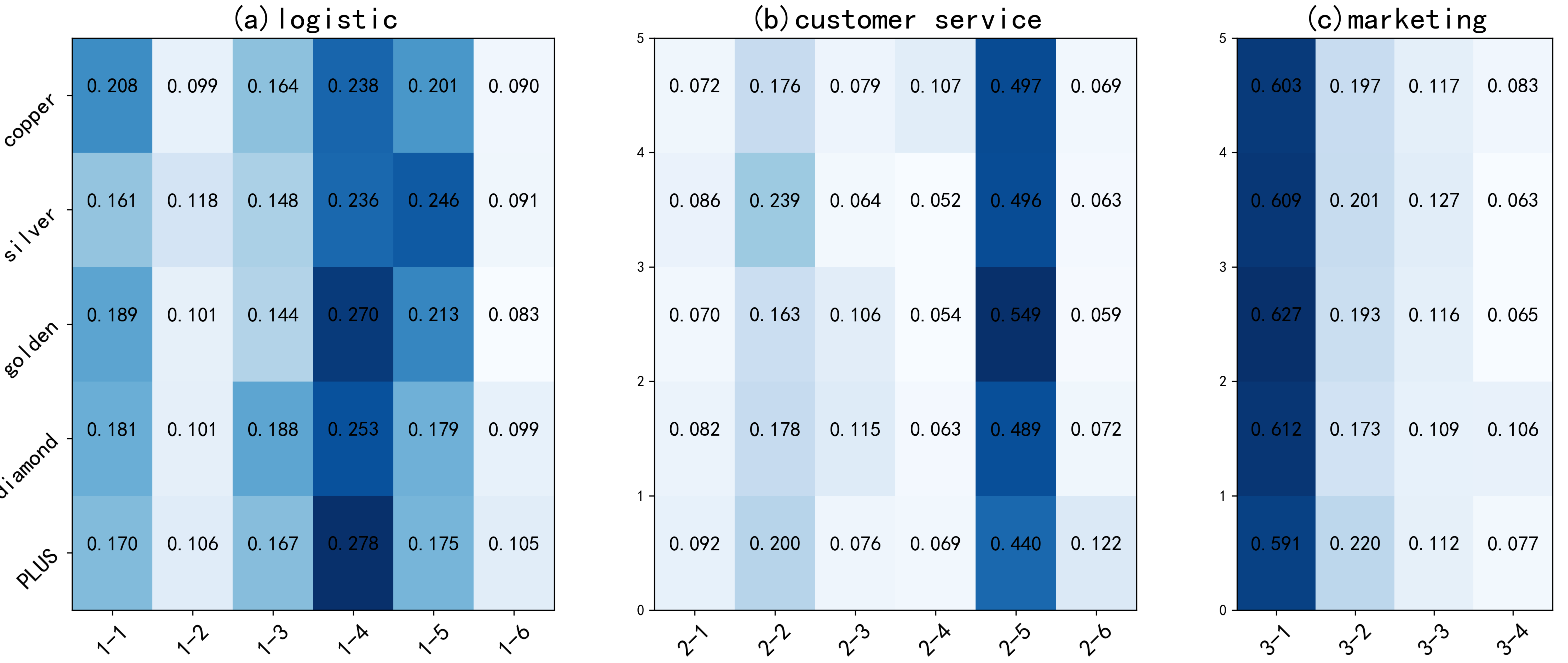}
\caption{\textbf{Proportions of detailed aspects of negative reviews for the three main reasons.} In each plot, the x-axis represents different aspects, as illustrated in Table 2, and the y-axis stands for different user levels. The horizontal sum of the patches in every subplot is 1.}
\label{Fig. 9}
\end{figure}

\paragraph*{}
To explore the deeper regulations or differences among users about how they regard negative reviewing, a division of only four main reasons is somehow insufficient. Therefore, we constructed an LDA model based on every main reason to further detect online consumers' detailed aspects for negative reviews. It should be noted that we did not construct an LDA model for the reasons of product defeats, considering that the content of a related review is always related to a product's function and design and is sensitive to the product category, lacking the possibility of stable clusters.  Therefore, we built three LDA models for complaints of logistics, customer service and marketing.
\paragraph*{}
We consider model perplexity and topic coherence together in setting the parameters of the LDA. Model perplexity can be characterized by a positive number that indicates the uncertainty level, while topic coherence is used to describe the similarity among different topics and can be characterized by Umass coherence \cite{Mimno:2011:OSC:2145432.2145462}, which is commonly a negative number indicating a better topic distribution through a smaller absolute value. According to this, we determined the criteria for the selection parameters as the multiplier of Umass coherence and perplexity, called the \textit{multiplier parameter}. Thus, a larger \textit{multiplier parameter} suggests a better performance of LDA. Moreover, it is worth mentioning that we prefer the one with a lower topic number to reduce the overfitting possibility, regarding two multipliers with close values. In line with this, we determine three models for the three main reasons, with topic numbers as 8, 9 and 7. The details of the parameter selection and topic words for every model can be seen in Fig. A3 and Table A2.
\paragraph*{}
According to the topic words (see Table A2) of each topic, we summarized complaints on logistics, customer service and marketing, as seen in Table 3. We also obtained the percentage of each aspect shown in Fig. \ref{Fig. 9}. More than 40\% of the logistics-related negative reviews complain about the low speed of delivery, followed by improper delivery timing with 18\% and packaging problems with 11\%. For customer service, complaints about shipping fees when returning goods come first, followed by rude customer service attitudes, delivery time and gift packaging negotiation issues. Thus, it can be proposed that the e-commerce platform should take more effective actions to enforce employee training, construction of logistics infrastructure and channels of information feedback. Moreover, an abnormal price increase or an abrupt price reduction after purchase serves as the most dominant issue in false advertising, suggesting that e-commerce platforms should pay more attention to the price monitoring and management aspects. These findings and inspirations indicate that understanding the behavioral patterns behind negative reviewing can indeed help to trace back to the existing deficiencies and spark new improvements.
\begin{table}[h]
\caption{Detailed aspects for complaints about logistics, customer service and marketing.}
\small
\centering
\begin{tabular}{L{1.5cm}|l|L{8cm}|L{1cm}}
\hline
Main reason & No. & Detailed aspects                                        & Topic no. \\
\hline
Logistics   & 1-1   & Poor attitude of courier service                 & 1         \\
            & 1-2   & Logistics status does not match the displayed one         & 2         \\
            & 1-3   & Slow logistics and slow delivery                          & 4,6       \\
            & 1-4   & Bad contact with courier                                  & 5         \\
            & 1-5   & Improper delivery timing                                  & 7,8       \\
            & 1-6   & Packaging problem                                         & 3         \\
            \hline
customer service & 2-1   & Problems about shipping fees when returning goods          & 1         \\
            & 2-2   & Rude attitudes of customer service                         & 2,5,6     \\
            & 2-3   & Return refund problem                                     & 3         \\
            & 2-4   & Invoice-related problem                                   & 4         \\
            & 2-5   & Delivery time and gift packaging negotiation issues       & 7,8       \\
            & 2-6   & Installation-related problem                              & 9         \\
            \hline
Marketing   & 3-1   & Abnormal price increase or abrupt price reduction after purchase & 1,2,4     \\
            & 3-2   & Lack of promised gifts, etc.                              & 3,7       \\
            & 3-3   & Invoice-related problem                                   & 5         \\
            & 3-4   & Overclaiming that does not match the real thing                  & 6         \\
\hline
\end{tabular}
\end{table}

\paragraph*{}
Additionally, we can further find discrepancy among user levels' preferences in Fig. \ref{Fig. 9}. For example, golden users have low tolerance of common problems, such as slow delivery and shipping fees when returning goods, while copper users are sensitive to consumer service attitudes and logistics speed. Even more inspiring, copper and golden users have more consistent preferences, while silver and PLUS users are more diversified in detailed aspects leading to negative reviewing. Though the main reasons that trigger negative reviews are level-independent, different preferences across levels in detailed aspects demonstrate that consumers of various levels post negative reviews for diverse reasons, implying that further examination of level-related discrimination is necessary.
\subsubsection*{Expression habits in reviewing}
\begin{figure}[h]
\centering
\includegraphics[height=7.0cm]{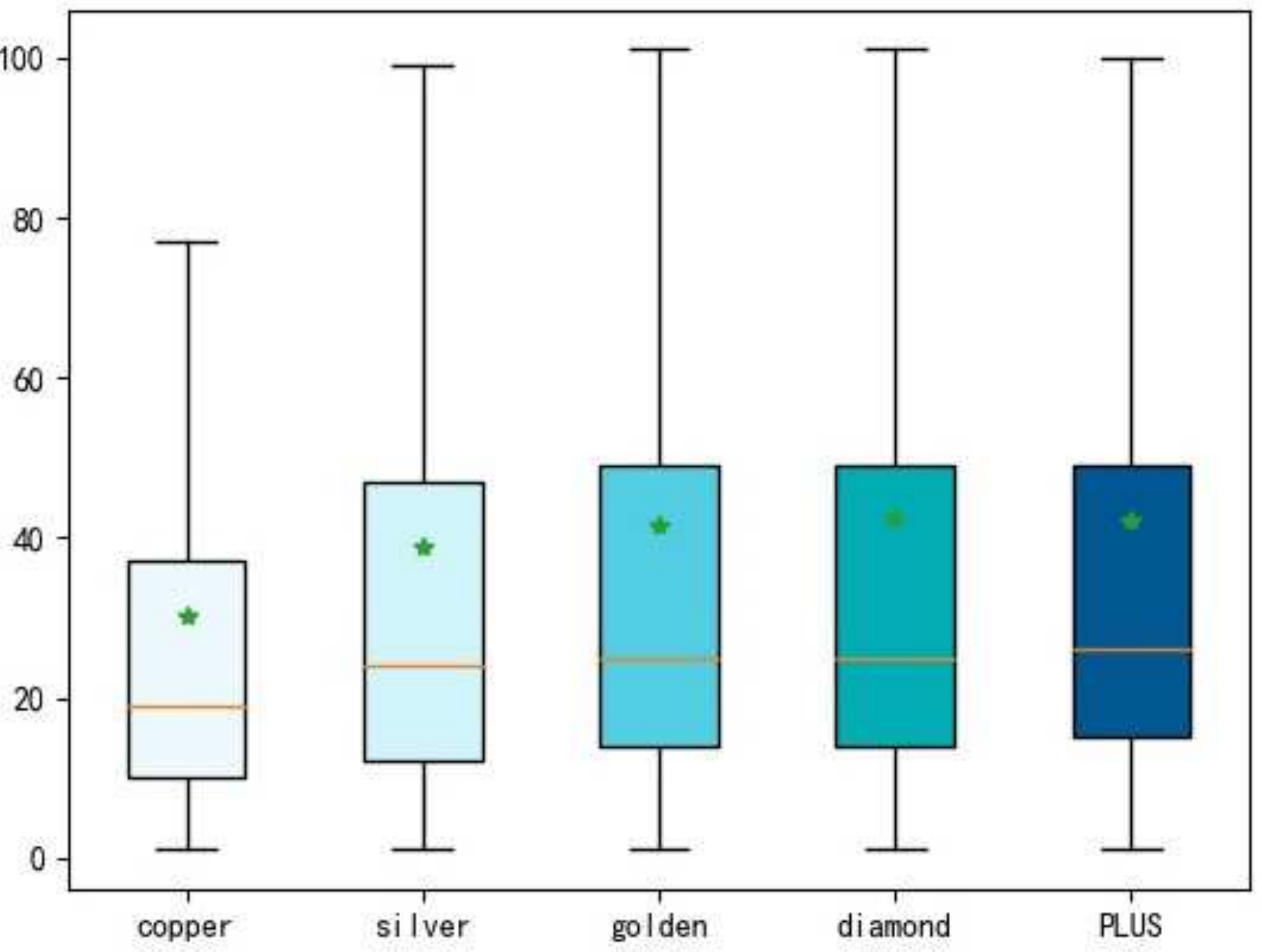}
\caption{\textbf{Rough distribution of negative review length of characters for the five user levels.} It is a no-fliers boxplot for negative review length in characters. The x-axis refers to different user levels, rising from left to right, and the y-axis refers to the review length in characters.}        
\label{Fig. 10}
\end{figure}

\paragraph*{}
The reasons for users posting negative reviews are directly reflected in a single sentence of reviews, while expression habits are explored through establishing a frequent connection in multiple sentences and are an externalization of buyer perception. In this paper, we attempted to characterize the expression habits of different user levels from the aspects of review length and semantic similarity.
\paragraph*{}
Review length is an indicator of review helpfulness, information diagnosticity \cite{Mudambi2010} and product demand \cite{Ghose:2012:DRS:2261184.2261192} and is an important factor for user centrality \cite{Tanase2015}. Here, we implement two ways to measure the review length, with the same outcomes. The first method is to measure the exact length of all characters within the text of the review, and the second is to count the number of effective words after cutting the text into a word list and filtering stopping words. Fig. \ref{Fig. 10} shows the distribution of negative review length for five user levels, in which the first method is employed. Considering most of the outliers come from the deliberate repetition of meaningless words and increase the difficulty of viewing and comparing among user levels, fliers are ignored here. It can be seen that with the increase in user level, the negative review length has an upward trend, and it is same as the performance of effective words of the second method (seen Fig. A4). And the upward trend is statistically significant in the one-way $t$-test. The growth of review length with levels implies that consumers of high levels post longer negative reviews and offer more reference and helpfulness for both consumers and sellers. From this view, the value of negative reviews from high-level users should be stressed in practice.

\begin{figure}[h]
\centering
\includegraphics[width=12cm]{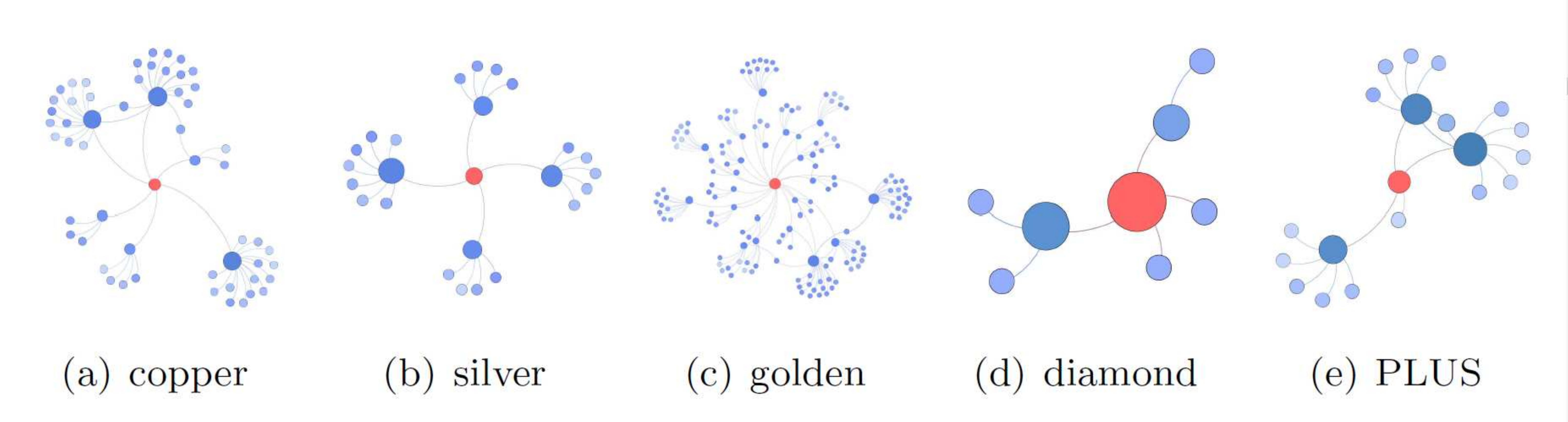}
\caption{\textbf{Semantic network of `match' for five user levels ($N=4$).} The red node in each network refers to the word `match', and the blue nodes are the words that are the top-four similar synonyms in each level. The sizes of the nodes refer to their degrees.}
\label{Fig. 11}
\end{figure}

\paragraph*{}
Differences in users' expression habits are reflected in the landscape of word similarity in contexts. Fig. \ref{Fig. 11} shows the semantic networks of the word `match' for five different user levels, where words connected to it are the top-four most similar synonyms of different contexts. As can be seen, users from different levels indeed demonstrate diversity in expression habits in terms of distinguishing network structures. According to this, we constructed word networks for each user level to examine the similarity of user language expression habits across user levels.

\begin{figure}[h]
\centering
\includegraphics[width=12.0cm]{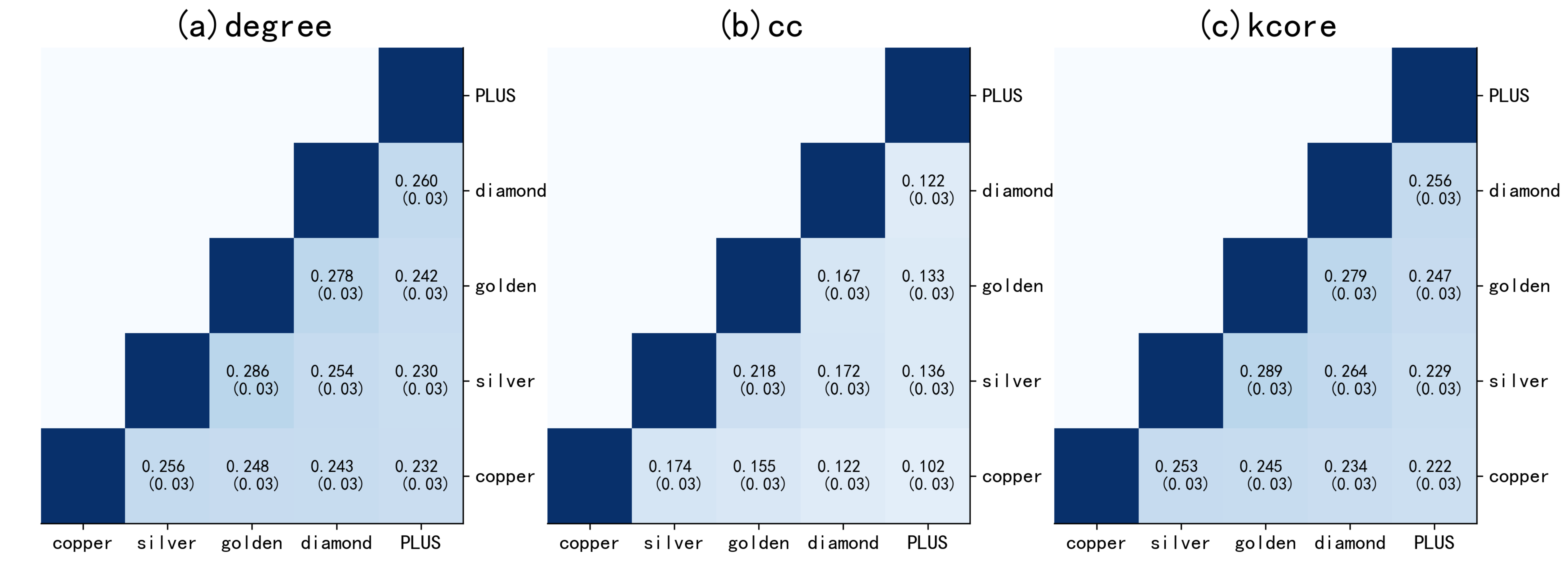}
\caption{\textbf{Similarity of expression habits across user levels.} The similarity is measured by the Kendall correlation coefficient of 500 randomly selected words, repeated 1000 times. The x-axis and y-axis represent five user levels, and the numbers in the patches refer to the average with the standard deviation in brackets behind it. Patches in the grid refer to the similarity between two user levels, with darker colors indicating greater similarity.}
\label{Fig. 12}
\end{figure}

\paragraph*{}
For the selection of $N$ in building word networks, the relatively giant connected component of $r_c$ is used to narrow the range. As $N$ increases, $r_c$ increases, implying that more words are connected to form a more complete landscape of semantic similarity. Until $r_c$ ends its highest rate of increase, it can be conjectured that the most representative and strongly connected words have been contained in the network. According to this, we set $N$ as 4 (see Fig. A5).
\paragraph*{}
After establishing five networks with $N=4$, the similarity of user language expression habits is measured by the Kendall correlation coefficient in a bootstrapping manner. Regarding the indicator, we focus on three different ones that characterize different features of the network: degrees measuring node connections, clustering coefficients measuring adjacent node connection density and k-core indexes, meaning node location within the network, e.g., the larger it is, the closer to the core. When implementing bootstrapping for a robustness check here, the repeating frequency is set as 1000 and the size as 500, i.e., 500 words are randomly selected from the common set for all user levels. Fig. \ref{Fig. 12} presents semantic similarity in the Kendall correlation among the five user levels. In line with expectations, the similarity at the diagonal is 1, meaning exactly the same. Furthermore, it can be observed easily that a user with such adjacent levels show greater similarity, implying more sharing in expression habits. The silver user and golden user obtain the greatest similarity of 0.22-0.28. On the other hand, regarding the absolute values of similarity, they are mostly in the range 0.1-0.3, suggesting that users of different levels are profoundly different in their expression of negative reviewing, which provides a further indication for an examination into emotions. In addition, the great differences among user levels are truly beyond our expectation, which indicates that if e-commerce platforms regard different users' expressions as the same, there will be a high possibility of misunderstanding users' reviews and then implementing improper actions. Note that the results from other settings of $N$ are consistent, and our above observations are robust (see Fig. A6).

\paragraph*{}
The exploration is aimed at perceptional discrimination in negative reviewing, examining causes beyond reviews and expression habits in reviews. Our investigations reliably demonstrate the existence of level-dependent patterns in negative reviewing. The conclusions and conjectures here provide the possibility of establishing perception models for precisely profiling different consumers. The profound differences in user levels also make the level-dependent strategies necessary in an e-commerce platform.

\subsection{Emotional patterns}
\begin{figure}[h]
\centering
\includegraphics[width=12cm]{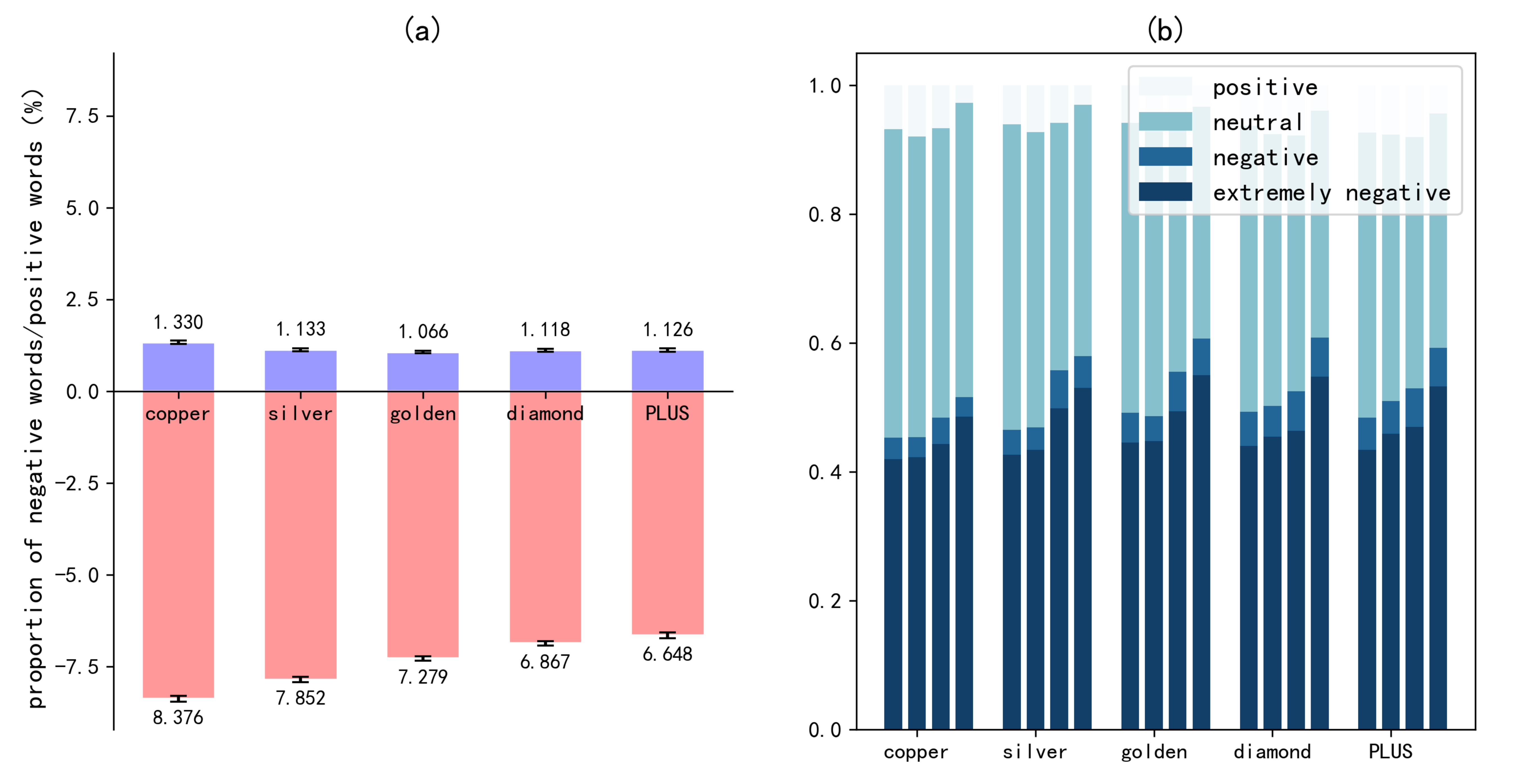}
\caption{\textbf{(a) Average proportions of positive and negative emotion words in negative reviews of 2017. (b) Distribution of user emotion from different user levels in four sectors.} (a) The bars on the upper side of the x-axis represent the percentage of positive emotion words, while those on the underside indicate that of negative ones. The five groups, from left to right, refer to user levels, that is, copper, silver, golden, diamond and PLUS. The absolute value of the y-axis means the percentage of emotional words. In addition, the error bar is calculated by the sample standard deviation. (b) A stacked histogram for emotion distribution of user levels. Bars of different colors represent the percentage of different degrees of emotion, which is divided into positive, neutral, negative and extremely negative according to $i_{polar}$. The x-axis refers to five user levels, and the four bars on a certain user level are for the sectors of Gifts\&Flowers, Clothing, Computers and Phone\&Accessories.}
\label{Fig. 13}
\end{figure}

\paragraph*{}
Similar to why consumers post negative reviews, the emotion features of consumers can also be extracted from the texts of reviews, along with scores that represent consumers' emotion baselines. However, different from content analysis of reasons for negative reviews, sentiment is an attribute for the entire review, which can directly influence how practitioners solve the complaints from consumers. Here, we regard negative reviews as a mixture of both negative and positive emotions, instead of only negativeness, and aim at the distribution of buyer emotion from different levels and its evolutionary patterns in both time and tendency.
\paragraph*{Distribution of user emotions}
Consumers from different levels may have differences in the distribution of emotion, such as the proportion of negative or positive sentiments and the extent of polarization. If they exist, some patterns can be used to guide sellers to employ more attention or take special actions regarding those with extremely negative feelings.
\paragraph*{} 
First, we calculate $r_{pos}$ and $r_{neg}$ to characterize the emotional degree of different user levels. Fig. \ref{Fig. 13}(a) shows the average percentage of positive and negative emotion words in negative reviews posted in 2017. As can be seen, the degree of negative emotion decreases as user levels increase, while there is no evident trend in positive emotion. Interestingly, golden users have the lowest positive words usage, and copper users deliver the highest degrees of both positiveness and negativeness. It can be summarized that consumers from upper levels are usually less emotional than those of lower levels. The results agree with our expectation, as consumers from upper levels have more experience in product selection and negotiation with sellers, leading to being less emotional but more rational. The results of the other years, from 2013 to 2016 (seen Fig. A7), are consistent with the stabilization of the user structure and the business model.

\paragraph*{}
The single index of $r_{pos}$ or $r_{neg}$ can reflect only the positive or negative degree of a review. To measure the polarization, we introduce the index $i_{polar}$ here to figure out the distribution of buyer emotion, as seen in Fig. \ref{Fig. 13}(b). It can be found, as expected, that the negative polarization dominates the negative reviewing for all consumers of all sectors. For sectors of Gifts\&Flowers, Clothing, and Phone\&Accessories, the negative and extremely negative polarization increases as the user levels increase, while for the sector of Computers, it shows an upward tendency first and then goes downward. For the upward tendency in the user level that is different from that shown in Fig. \ref{Fig. 13}(a), it can be interpreted as the different compositions of the two indexes. The index in Fig. \ref{Fig. 13}(b) omits words that are neither positive nor negative in the sentiment dictionary. Therefore, the different results can be explained by considering that consumers with lower levels tend to use more negative words than upper-level users; however, in negative reviews from upper-level users, positive words co-occur with negative words less, leading $i_{polar}$ to be equal to or near -1. Through these findings, we can conclude that upper-level users are less emotional than lower-level users, but the emotion expression in negative reviewing is more condensed and concentrated.

\paragraph*{}
To further probe the differences in the usage of emotional words across five user levels, we calculate the possibility of appearing in one negative review for each word $w$ in the sentiment dictionary, which is defined as $f_w=n_w/n_r$, where $n_r$ is the number of all negative reviews posted by users of a certain level, and $n_w$ is $w$'s occurrences within these reviews. Moreover, for each word, we also calculate the variance in occurring possibilities across the five user levels and filter out words with a large variance to infer the differences in emotional preferences, as seen in Table 3. Interestingly, it can be concluded that users from lower levels place more importance on service attitudes and whether the product matches what they are in marketing, while upper-level users tend to have diversified shopping demands and preferences and emphasize the overall experience. The diverse demands of upper-level users, along with their more condensed review content, can lead to a less biased narration. This observation is consistent with the findings from the perception of the causes leading to negative reviewing.

\begin{table}[h]
\caption{Emotional words with different trends as the user level increases.}
\scriptsize
\centering
\begin{tabular}{p{3.5cm}|p{1cm}|p{3.5cm}|p{1cm}}
\hline
\multicolumn{4}{c}{\textbf{Emotional words}}\\
\hline
\textbf{fre. of usage increasing as the user level increases (copper, silver, golden, diamond, PLUS)}                                                        & \textbf{aspect }                            & \textbf{fre. of usage decreasing as the user level increases (copper, silver, golden, diamond, PLUS)}                   & \textbf{aspect}                         \\
\hline
'Authentic','Original',   'brand','word-of-mouth','Quality assurance'                                 & brand \&word-of-mouth               & \multirow{5}{3.5cm}[-80pt]{'attitude in service', 'suitable','match','retread','Be deceived'}& \multirow{5}{1cm}[-80pt]{used product\& service attitude} \\
\cline{1-2}
'cheap price','on sales','Gift','Cost effective','Good value for money','difference of prices'  & price                              &~ &~                                 \\
\cline{1-2}
'smooth','fluently','material', 'stable/stability','clear/clearness', 'texture','clean/cleanness'   & feeling of usage\& product texture &~&~                                 \\
\cline{1-2}
'solve the problem',  'professional','responsibility', 'patience/patient','protection', 'user-friendly' & quality of service                 &~& ~                                \\
\cline{1-2}
'integrity','sincere','shop bully','rights protection','fraud/fudge'                               & integrity                         &~&  ~                               \\
\hline
\end{tabular}
\label{Table 3}
\end{table}

\paragraph*{Emotion evolution over time}
Online reviewing might occur continuously, such as posting multiple times in the same day. To verify this, we test the continuous probability in one day for the PP and RC time of reviews from identified consumers. The results show that the continuity for RC time is stronger than that of PP time in three sectors, and the remaining one has two numbers that are very close, indicating that compared with purchase time, reviewing time is more continuous (see Fig. A8). In line with this, a natural question is to consider how consumers' emotions evolve in the continuous reviewing, which makes the examination of the emotion sequence necessary.

\begin{figure}[H]
\centering
\includegraphics[width=12cm]{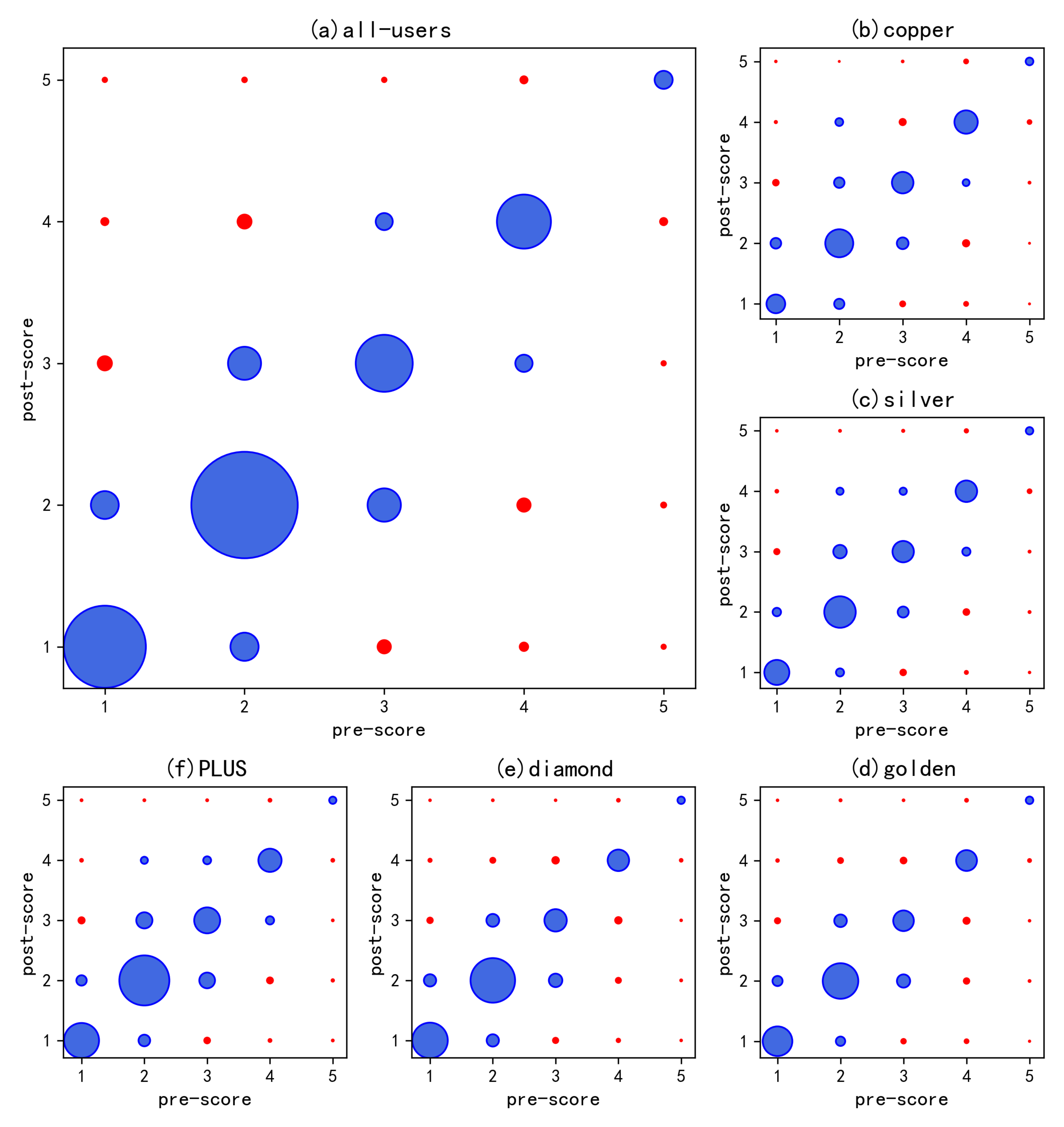}
\caption{\textbf{Normalized continuous probability for every score pair with a threshold of an hour.} The x-axis and y-axis are the scores of the previous reviews and the postreviews, respectively. For every bubble, we normalized it by dividing the initial probability by the corresponding postscore's proportion, the colored bubbles with a value larger than one are blue, and others are red. Therefore, the size, along with the color of the bubbles, represents the extent of the influence that comes from the emotional basis of previous reviews.}
\label{Fig. 14}
\end{figure}

\paragraph*{}
In addition to sentiment indexes, the review score is also a quantitative indicator of the emotional attitude and can be employed to reflect the evolution of emotion. For every identified user, we sort the scores of continuous reviewing in a time-ascending order and then cut sequential scores into sequences in which the intervals of adjacent scores should be smaller than a specified threshold, e.g., 0.5 or one hour. The continuous probability of every pair of adjacent scores is calculated to determine whether the postscore will be influenced by the former one, providing evidence for the emotional evolution over time. For example, a score pair can be defined as (1,2), which means that the previous one is score one and that the post one is score two. Meanwhile, to eliminate the influence of uneven occurrences of different scores, the proportion of five scores in all reviews is implemented to normalize the continuous probability. Fig. \ref{Fig. 14} shows the continuous probabilities after the normalization. From the figure, we find that all users demonstrate a momentum in negative emotion, which is shown by the diagonal bubbles that are larger than the others; moreover, the farther away from the diagonal, the smaller the bubbles are. In addition, comparing Figs. \ref{Fig. 14}(b)-(f) suggests that users from upper levels have a stronger momentum in negative emotion than those of lower levels, as the size of the diagonal bubbles become larger as the user level increases. For every threshold in [0.5,1,2,24], we obtained the same conclusion, as a promising indicator of robustness. In contrast with the conclusion above that the upper-level users are less emotional, a more intensive momentum in negative emotion indicates that although upper-level users have a richer set of experience and more stable emotions, once they are faced with an unsatisfactory experience, it will cost more for sellers to comfort them and alleviate the negative outcomes.

\begin{figure}[!h]
\centering
\includegraphics[width=11cm]{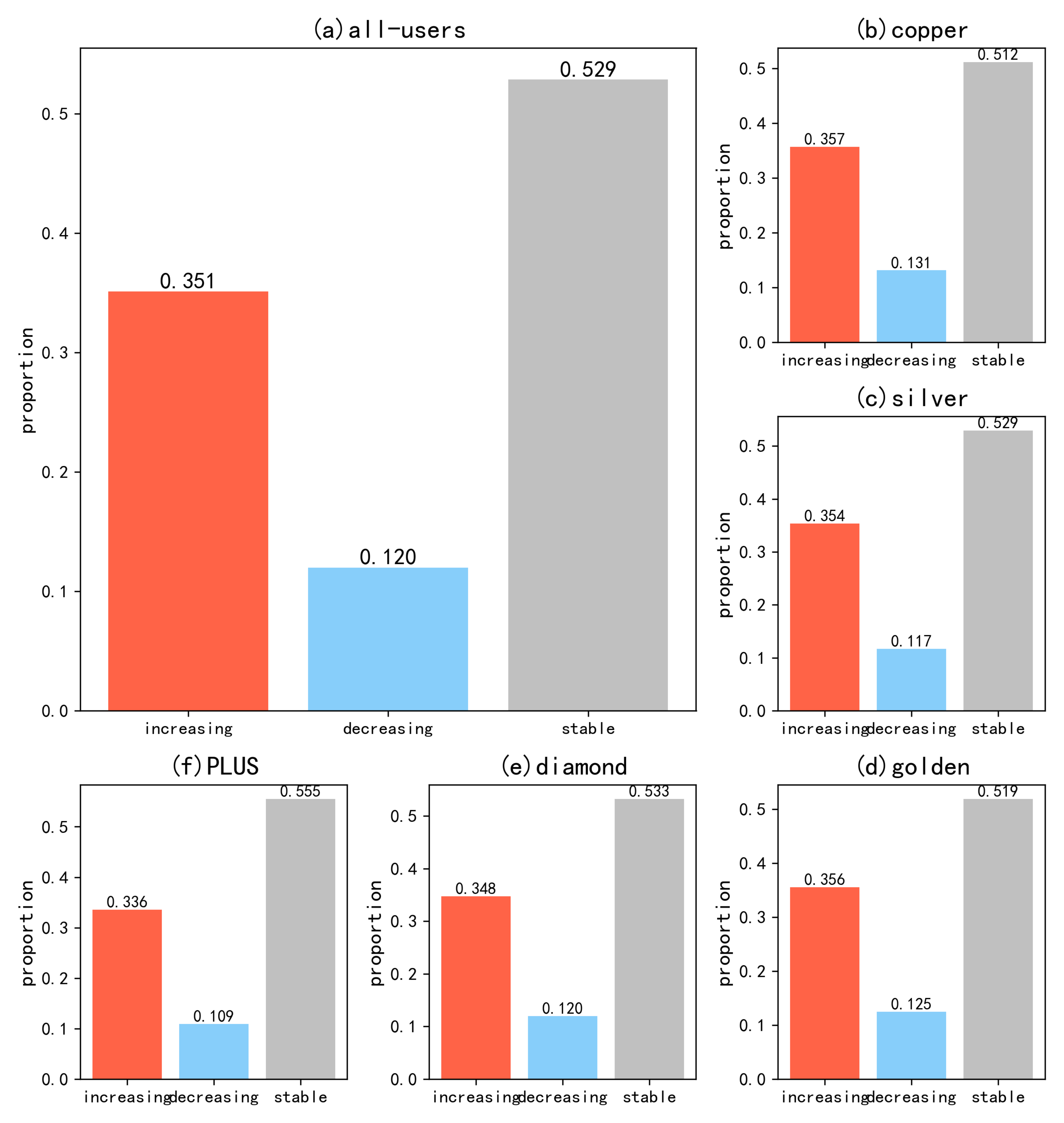}
\caption{\textbf{Proportions of emotion shifts in negative reviews of all users and user levels, taking the Computer sector as an example.} The x-axis shows three types of emotion shifts. The height of different colors bars refers to the percentage of a certain kind of emotion shift in all after-use comments with the initial score equal to 1. The results from other sectors are consistent.}
\label{Fig. 15}
\end{figure}

\begin{figure}[H]
\centering
\includegraphics[width=11cm]{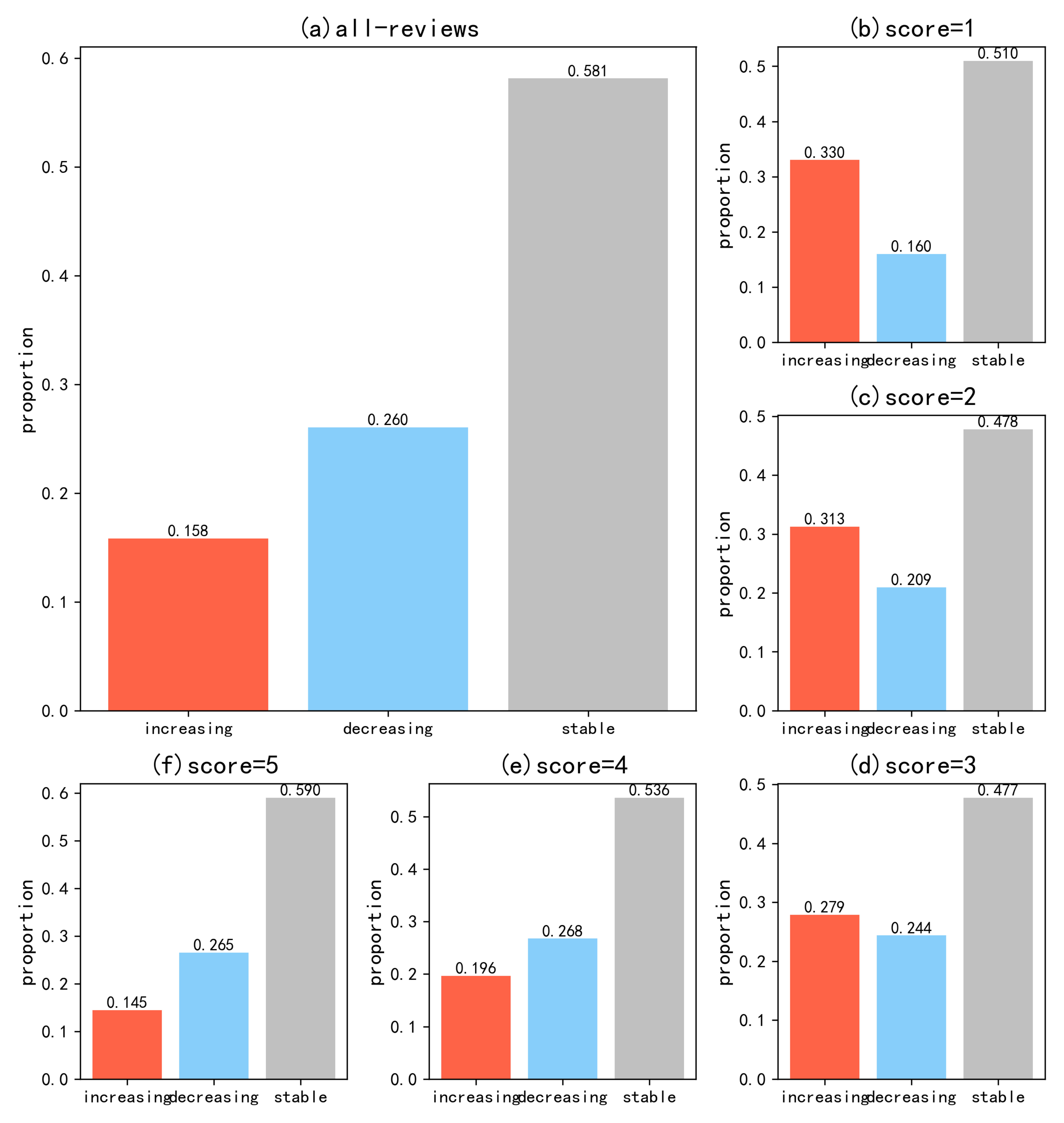}
\caption{\textbf{Proportion of emotion shifts in all reviews and all reviews with different scores, taking the Computer sector as an example.} The x-axis shows the three types of emotion shifts defined above. The height of different colors bars refers to the percentage of a certain kind of emotion shift in all after-use comments. Note that the results from other sectors are the same.}
\label{Fig. 16}
\end{figure}

\paragraph*{Tendency of emotion shifting}
Emotion toward a certain objective is not static and will change over time due to external factors. To explore how online consumers' emotion shifts over time, we focus on reviews with after-use comments, which are posted after the initial review, usually separated by a certain time of product usage. Posted after-use comments, compared with the initial one, often contain some shifts regarding usage experience and the resulting attitudes and emotions. According to the polarization $i_{polar}$ pair for the initial review and after-use comment, abbreviated as ($i_{polar-initial}$,$i_{polar-afteruse}$), respectively, we defined three directions of mood shifting, that is, increasing with the pair as (-,+), (0,+) or (-,0), implying a shift toward more positive emotion, decreasing as (+,0 ), (+,-) or (0,-), implying a shift toward more negative emotion, and stable as (0,0), (+,+) or (-,-), which implies no conversion. To reveal the exact patterns of emotion shifts beyond negative reviewing, we examine the occurrence of difference shift directions and the time needs for shifts across both user levels and review scores.
\paragraph*{}
Regarding the proportion of three shift directions, as shown in Fig. \ref{Fig. 15}, more than 35\% of consumers who post negative reviews obtain emotional improvement, which can be an indicator of how effective the action was that was taken after the poor rates. Moreover, there is a slight downward trend of the increasing situation as the user level increases, suggesting that it is more difficult to change upper-level users from dissatisfied to satisfied. Then, in Fig. \ref{Fig. 16}, the scores of initial reviews are taken into consideration. It can be observed that the gap first narrows between percentages of emotion increasing and decreasing and then increases in the opposite direction. This trend proved that the saturation effect occurred in the emotion shifts, which is described as always being much easier for emotion to develop in reverse rather than to become stronger in the initial direction. This result further implies that the shift toward the opposite direction will easily trigger the posting of after-use comments.

\paragraph*{}
The interval between an initial review and the after-use comment is calculated to measure the lag needed by the emotion shift, with the resolution of an hour. In Fig. A9, we provide a different perspective from the user level. It can be seen that for all negative reviews with after-use comments, lags become longer as the user level increases, except that a slight fluctuation occurs with PLUS users. Furthermore, the same phenomenon can be seen only in negative reviews with an increasing emotion shift. Therefore, it is conjectured that the average interval becomes larger as the user level increases, together with the lags for emotion improving. This suggests that users of upper levels may need a longer time to change their negative impression of products, which agrees well with our conclusion about upper-level users more intensive momentum in negative emotion.

\paragraph*{}
In this section of sentiment patterns, we focus on the polarization, evolution features and shifting emotions in negative reviewing and determine that lower-level users tend to have a more intensive negative emotion, while upper-level users demonstrate a more intensive emotion momentum when faced with unsatisfactory shopping experience. Moreover, patterns in emotion shifts prove again the conjecture about upper users' negative momentum and the existence of the saturation effect in emotion. These conclusions or conjectures can provide e-commerce practitioners with the potential to mitigate or avoid the negative word-of-mouth from breeding and spreading negative emotions.

\section{Discussions}
\paragraph*{}
Online reviews and consumers' behavior on e-commerce platforms has been widely studied; however, with respect to negative reviews that can be regarded as a kind of additional reference information for consumers and as a focus on performance improvement for sellers, the behavioral patterns that underlie them have not been explored in a comprehensive manner in the literature. Moreover, existing research aimed at online consumer behavior does not distinguish online users from different levels and only pays attention to a few aspects of features, which is not sufficient to obtain a deep and systemic understanding of consumers' characteristics. 
\paragraph*{}
This paper attempts to provide a systemic understanding of negative reviewing from temporal, perceptional and emotional aspects. It utilizes a multisector and multibrand dataset from JD.com, the largest B2C platform in China, and implements various methods of semantic analysis and statistics to offer a first empirical experiment based on a Chinese e-commerce dataset with more than several million pieces of data. Our main findings are the following:
\paragraph*{}
(1) Circadian rhythms exist with regard to negative reviews after the buying, which are related to the time consistency of people's daily activities. (2) The similarity of different users' expression habits from adjacent levels is significantly greater than that of other level pairs, and users with lower levels are more sensitive to prices and a seller's deceitful acts concerning pricing or to rude attitudes of sellers, while demands of users at higher levels are more varied and exhaustive. (3) In the emotion dimension, users at lower levels experience a more intensive expression of negativeness but with a lower emotional momentum in negativeness.
\paragraph*{}
Our study has implications for both academics and practitioners. For academics, we contribute to an enhanced understanding of online consumers' negative reviews, from the perspectives of temporal patterns, what kind of experience they tend to regard as a failure, and emotional patterns regarding how they express and how this expression changes. Although there have been many recent studies that examined features of online reviews and negative reviews, to our knowledge, there is a lack of research aimed at online negative reviewing. In addition to insights into behavioral patterns of negative reviewing, the richness of our empirical setting provides a possibility for future research to dig deeply into not only online negative reviewing but also other online behaviors within buying. In addition, some unexpected findings shed light on the diversity of users' behavior, such as the relationship between purchase time and reviewing time, the great differences among users' expressions, and the different amounts of emotion momentum, concepts previously examined in a limited way.
\paragraph*{}
For practitioners, some findings in this article are important because they offer guidelines or tools to determine a platform's or a seller's problems or disadvantages and then to devise targeted management strategies to improve a corporation's performance. The results about characteristics of different user levels underscore the necessity of implementing a distinguished strategy toward different user groups. For example, preferences of negative reasons suggest that managers should provide more purchase and selection directions for junior users with less experience but more extremely negative expressions, which will harm a platform's reputation and provide the personalization to review display on platforms. In addition, according to momentum in negative emotion, even though upper-level users employ less negative word-of-mouth, as long as they experience unsatisfactory shopping, the platform should pay more in costs or measures to comfort and alleviate their negative mood. Additionally, periodic intervals indicate that in addition to personal preferences on products, the time line of regular activities can also be modeled to help user profiles and precise marketing. These proactive strategies could boost sales, reduce negative word-of-mouth and increase consumer satisfaction.
\section{Conclusion}
\paragraph*{}
In this article, we have studied online consumers' behavior of negative reviewing in a systemic and user-distinguished perspective. Our analysis dimensions range from temporal and perceptional to emotional aspects and, to some extent, offer a comprehensive understanding of the online consumer complaint behavior compared to prior literature that mostly lacks richness of analysis dimensions. We found circadian rhythms in negative reviewing. Moreover, detailed causes leading to negative reviews and expression habits across user levels are exactly profiled. In addition, users from lower levels express more intensive negativeness, but those from upper levels demonstrate a stronger momentum in negative emotions.
\paragraph*{}
In the present work, the geographic location of consumers is not considered due to the shortage of relevant information in our dataset. In the future, related research can pay more attention to geographical features in negative reviews to help provide insights into geographical differences.

\section*{Competing interests}
  The authors declare that they have no competing interests.


\section*{Acknowledgements}
This work was supported by NSFC (Grant No. 71871006).


\newpage
\section*{Supplementary Information}
\setcounter{table}{0}
\renewcommand{\thetable}{A\arabic{table}}
\setcounter{figure}{0}
\renewcommand{\thefigure}{A\arabic{figure}}

\begin{figure}[ht]
\centering
\includegraphics[width=12cm]{fig1.pdf}
\centering
\caption{\textbf{Overview of dataset.} (a) The stacked histogram of different scores' proportions for four sectors. (b) Log-log distribution of usefulCount for every review, taking the sector Computers as example. (c) The stacked histogram of identified buyers for five user levels, distinguishing four sectors. (d) The stacked histogram of different scores' proportion for five user levels, distinguishing four sectors.} 
\end{figure}
\newpage
\begin{table}[H]
\caption{A brief description of the attributes of our dataset.}
\scriptsize
\centering
\begin{tabular}{p{2.3cm}|p{1.8cm}|p{5.2cm}|p{4.6cm}}
\hline
Name                        & Type of Data                                                                                                                 & Description                                                                                                                                                       & Snippet                                                                                                                         \\
\hline
Review                      & text                                                                                                                         & Text that consumers post on an e-commerce platform that expresses their attitude and evaluation regarding goods always has a numerical score or photos about goods. &                           
 ``The outer casing is severely deformed, the cooling effect is poor, and the noise is large.''\\
 \hline
Score of Review             & int                                                                                                                          & A numerical score on goods from buyers, a representative of consumer's integral evaluation, with the lowest score of one for a poor experience to the highest score of five for good one.   & A review about clothing that receives a score of five is ``It fits me well, and the seller is really patient.''                                  \\
\hline
Negative Review             & text                                                                                                                         & The reviews that receive the lowest score of 1 often represent an unpleasant purchase experience.                                                                                    & ``The outer casing is severely deformed, the cooling effect is poor, and the noise is large.'' \\
\hline
Sector                    & text(with a unique code)                                                                                                     & A product's category on an e-commerce platform, formulated according to the function or features of goods. It is usually retrieved from the platform itself.                & Consumer Electronics, Women's Clothing, Men's Clothing , Sports \& Outdoors, etc. (JD platform)                                \\
\hline
Subsector                 & text (with a unique code)                                                                                                     & A further subdivision from the sector.                                                                                                                                & sector Computers, with subsector: packaging machine, CPU, SSD, laptop, UPS, U disk, etc.              \\
\hline
Time span                   & date-date                                                                                                                    & The time scale of the valid reviews in our dataset from JD for every category.                                                                                    & 2008.11.02-2018.03.22 (sector Computers)                                                          \\
\hline
UserLevel                   & text                                                                                                                         & The level identification of users in the e-commerce platform. It can be divided into general user(the level grows with more purchases) and paid user.                    & In JD platform, General user: Registered /Copper / Silver / Golden / Diamond User; Paid User: PLUS User                              \\
\hline
Anonymous /identified user & Boolean                                                                                                                      & A Boolean flag describing whether the whole nickname of a reviewer can be seen in our dataset. If True, then Identified User; else, Anonymous User.                     & An the anonymous user in our dataset contains ``***''; An identified user is buaa\_jduser.                                           \\
\hline
UsefulCount                 & int                                                                                                                          & A numerical count of how many helpful votes a review received from other users until the time we collected the data from JD platform.                                  & The review ``The outer casing is severely deformed, the cooling effect is poor, and the noise is large.'', received five usefulCount.  \\
\hline
ReplyCount                  & int                                                                                                                          & A numerical count of how many replies a review received from the seller behind the review that he/she posted until the time we collected the data from the JD platform.             & The review ``The outer casing is severely deformed, the cooling effect is poor, and the noise is large'' received five of replyCount.   \\
\hline
After-user review           & text                                                                                                                         & The reviews that users post after the first review on certain product that often contain information or attitude that is updated.                                            & ``After using it for a while, I feel that my skin is really getting better.''  \\
\hline
User Client                 & text(with a unique code)                                                                                                     & Client type used when a user logs in,                                                                                                                               & e.g., iPhone, Android, WeChat, iPad, Web Page, etc.                                                                               \\
\hline
\end{tabular}
\label{Table A1}
\end{table}

\begin{figure}[H]
\centering
\includegraphics[width=12cm]{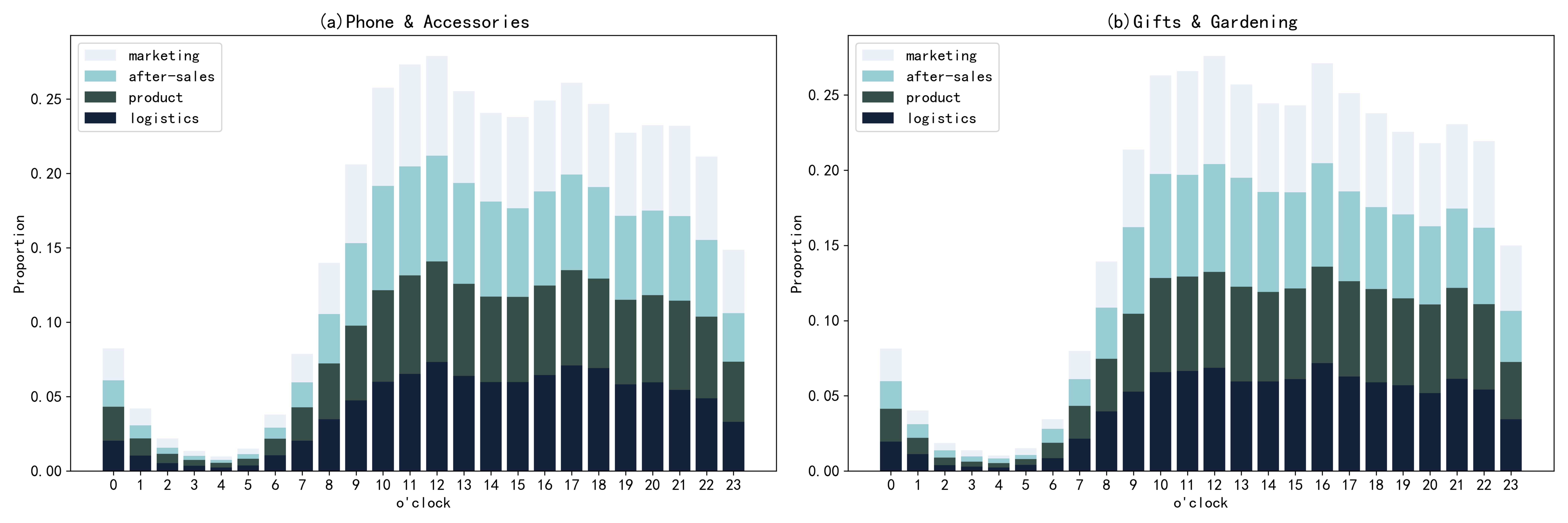}
\caption{\textbf{Proportions of main reasons for negative reviews on an hourly scale for the other two sectors.}
      (a) is a stacked histogram for sector Phone\&Accessories and (b) is for Gifts\&Flowers. The x-axis is  hours in a day, from 0 to 23, and the y-axis represents the proportion at the corresponding hour. Compared with the two sectors in Fig. 11, it can be concluded that Phone\&Accessories are products with strong personal features such as Clothing, while peaks in the sector Gifts\&Flowers often appear at 10-12, 16 and 21-22 o'clock, which is often delivery time.}
\end{figure}
\newpage
\begin{figure}[H]
\centering
\includegraphics[width=12cm]{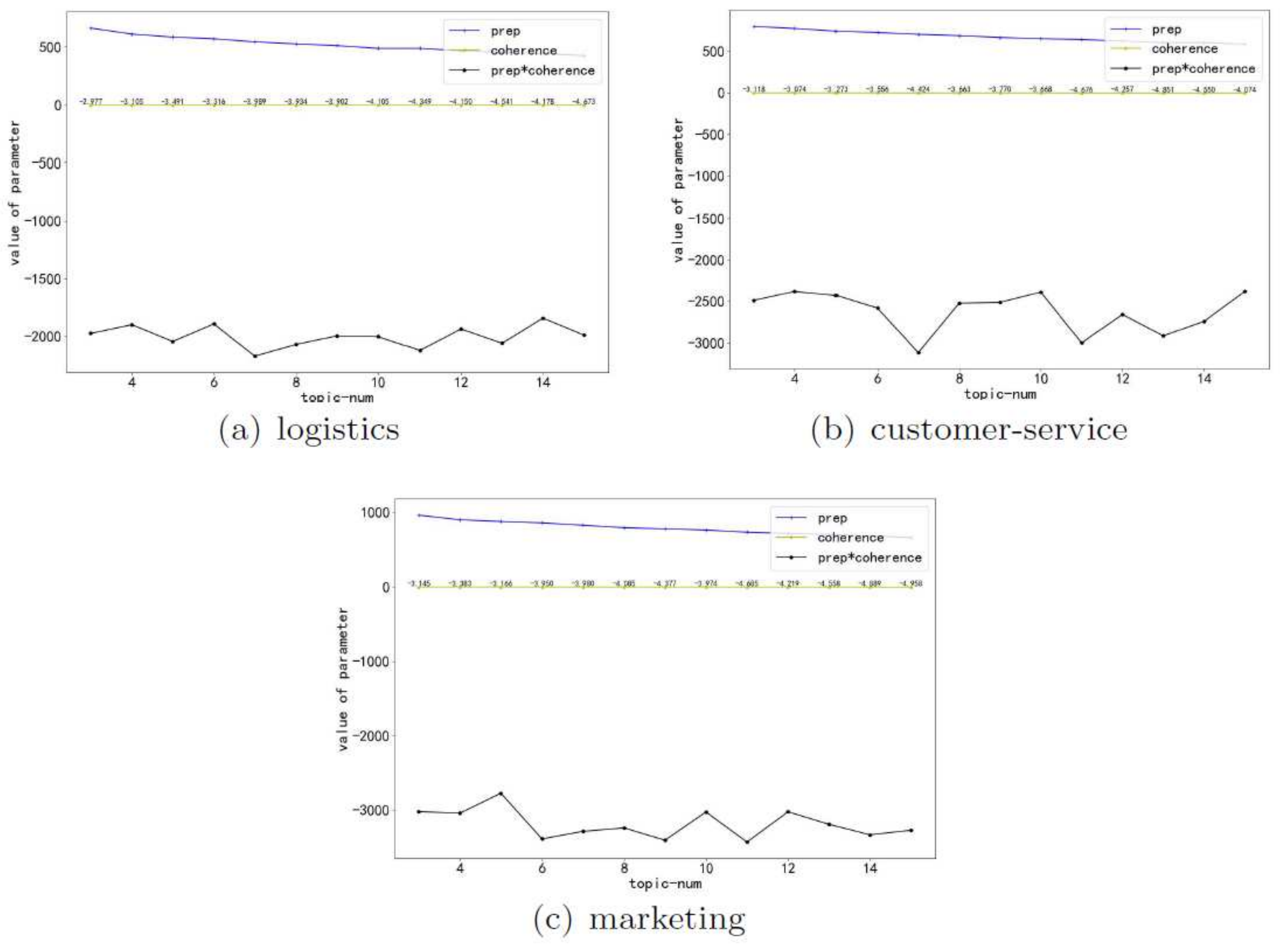}
\centering
\caption{\textbf{Parameter selection.}
  (a), (b) and (c) are curves for the main reasons of logistics, customer-service and marketing, respectively. The x-axis corresponds to the number of topic for model training, and the y-axis represents the value of multipliers. It is worth mentioning that this curve is not the only basis of parameter selection. The topic content and topic number are also considered.}
\end{figure}

\newpage
\begin{figure}[H]
\centering
\includegraphics[height=7.0cm]{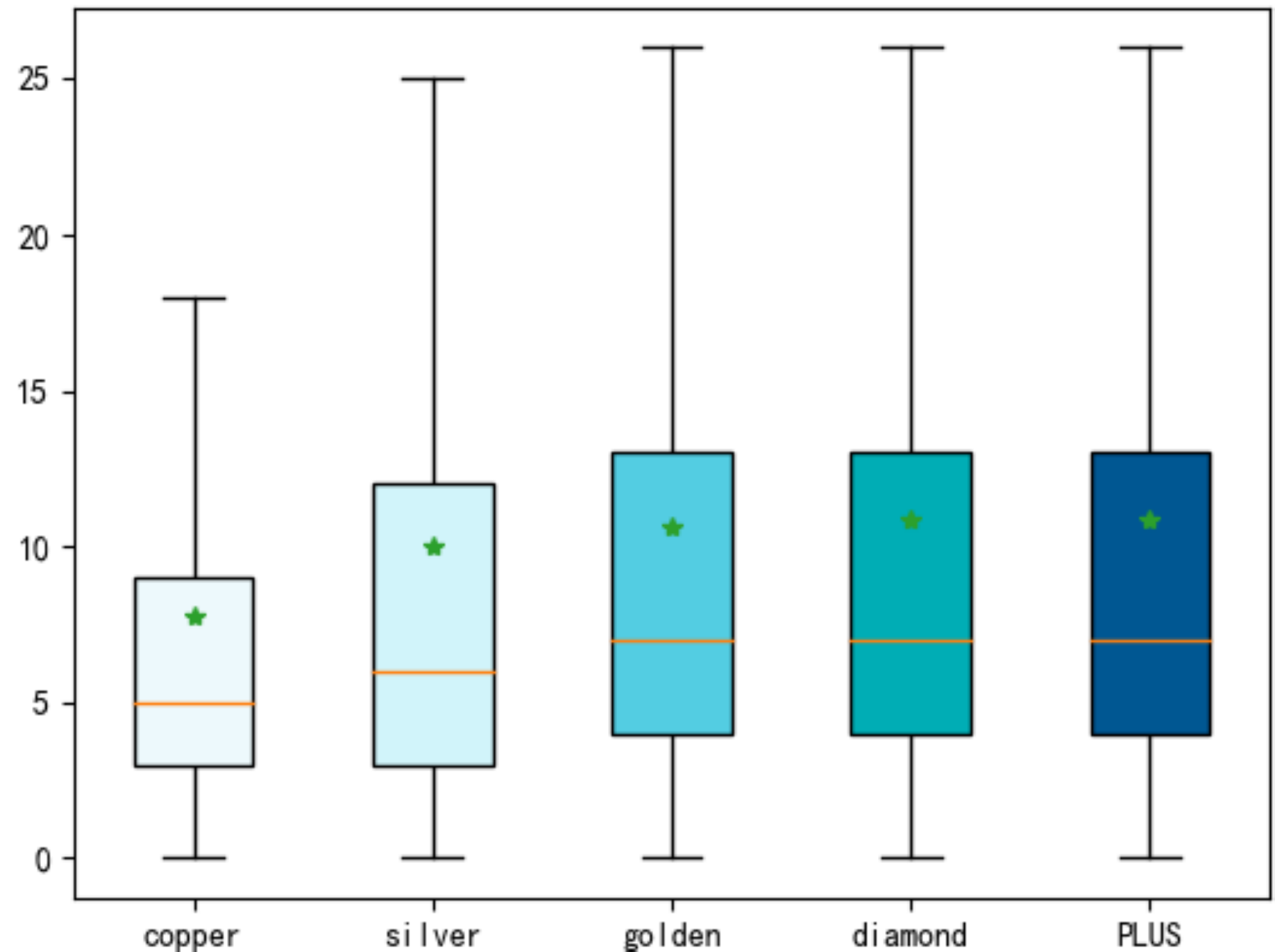}
\caption{\textbf{Rough distribution of negative reviews' length of effective words for five user levels.}
      It is a no-fliers boxplot for negative reviews' length of effective words. The x-axis refers to different user levels rising from left to right, and the y-axis refers to reviews' length in characters. And the growth of review with user levels is statistically significant in the one-way $t$-test. Compared to Fig. 10, it obtains the same conclusion about the review length of user levels.}
\end{figure}
\newpage

\begin{figure}[H]
\centering
\includegraphics[width=10cm]{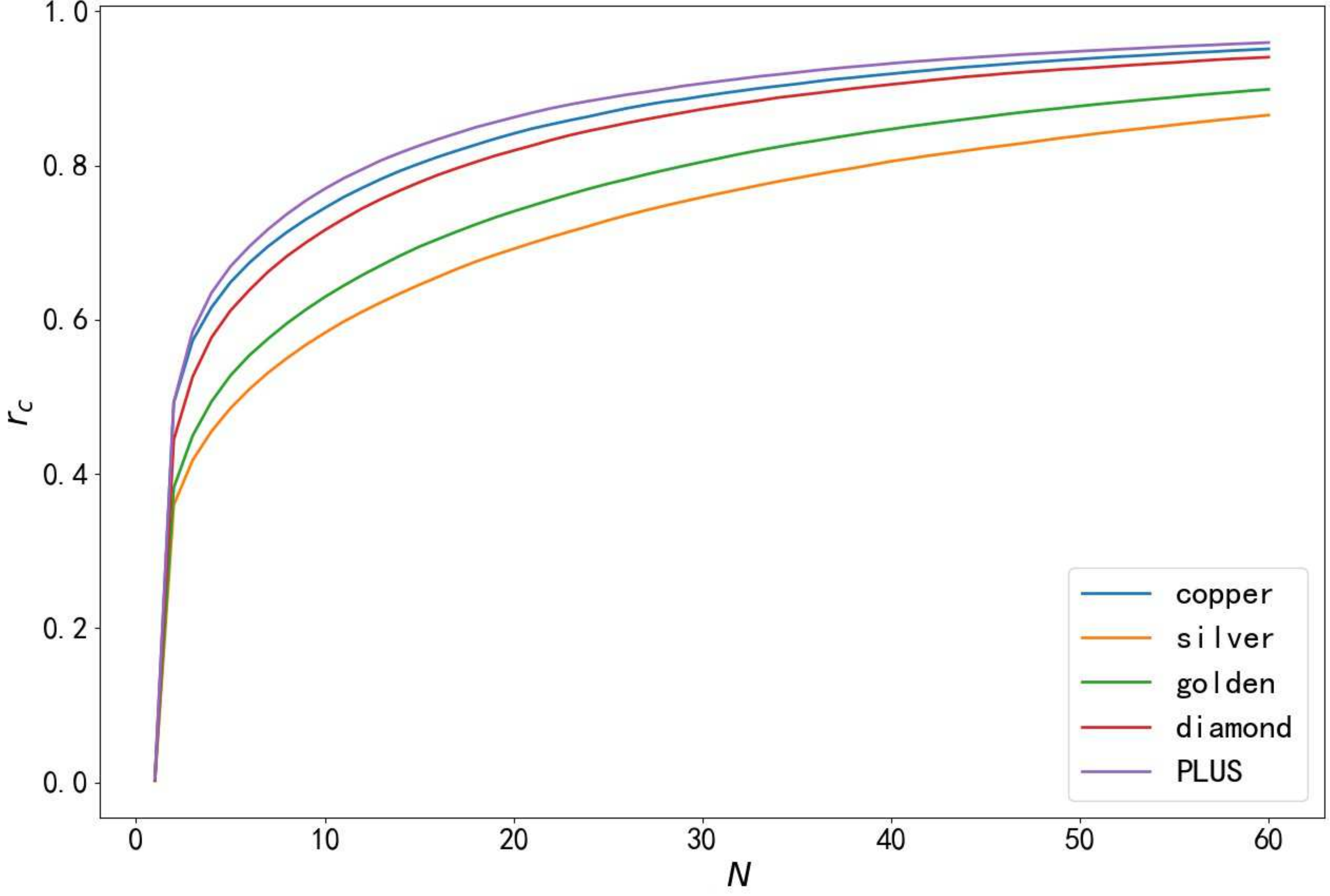}
\caption{\textbf{$r_c$ curve for all five user levels.}
      The x-axis shows $N$, the number of the most similar words chosen to form the network, and the y-axis shows the value of $r_c$. Five curves with different colors represent five user levels. For the selection of $N$, we aim at the one with or near the largest second derivative on the curve, which refers to the most effective or representative part that is contained in the network. Therefore, $N=4$ is determined, with $N=5,6,7$ to prove the robustness.}
\label{Fig. A5}
\end{figure}

\begin{figure}[H]
\centering
\includegraphics[width=11cm]{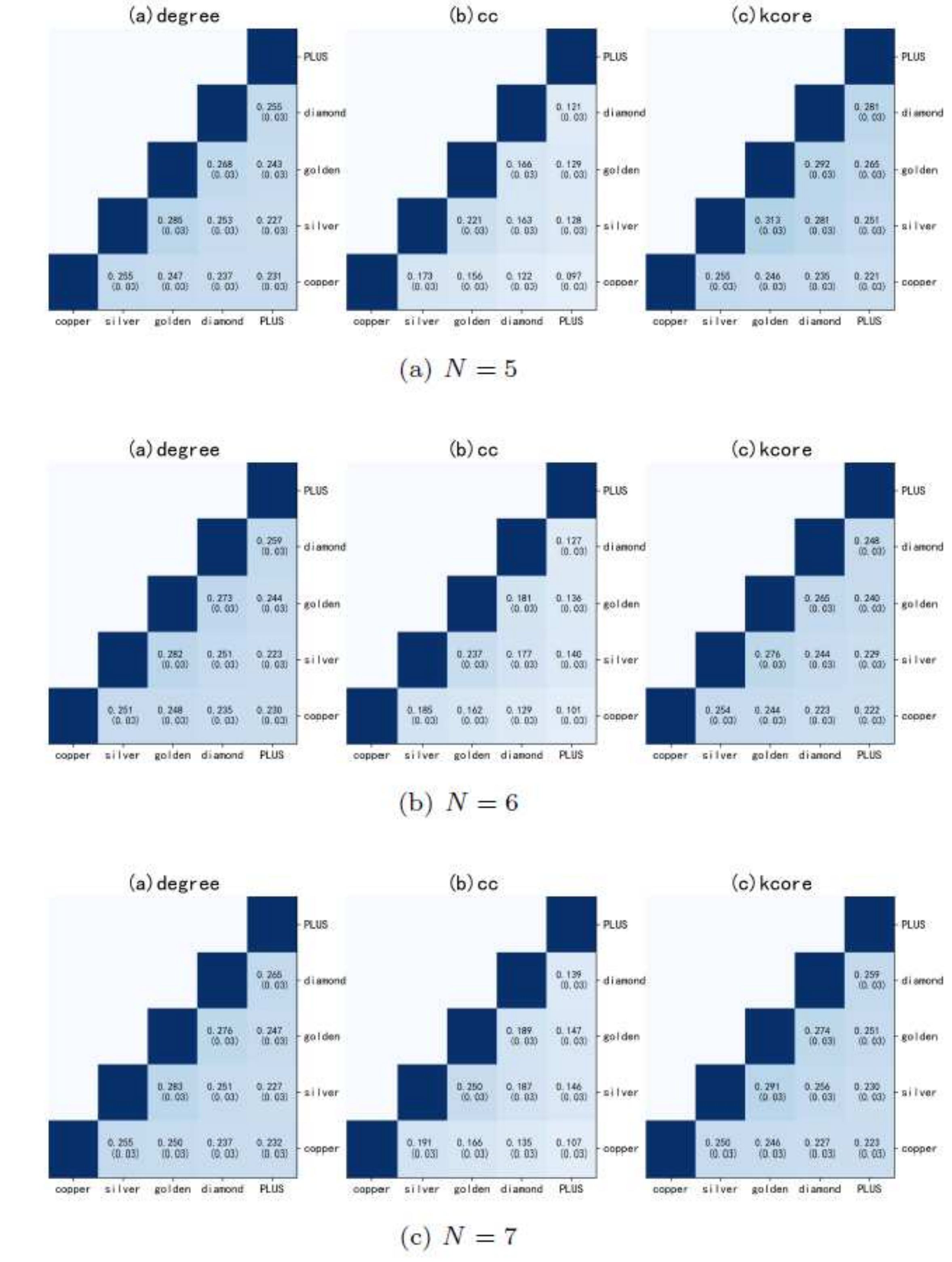}
\caption{\textbf{Semantic similarity among five user levels in the bootstrapping method ($N=5,6,7$).}
      The similarity is measured through the Kendall correlation coefficient of 500 randomly selected words that are repeated 1000 times. The x-axis and y-axis represent five user levels, and the numbers in the patches refer to the average with the standard deviation in brackets behind it. Patches in the grid refer to the similarity between two user levels, with darker colors indicating greater similarity. In addition, the outcomes here illustrate the reliability of the semantic similarity among user levels.}
\end{figure}
\newpage

\begin{figure}[H]
\centering
\includegraphics[width=10cm]{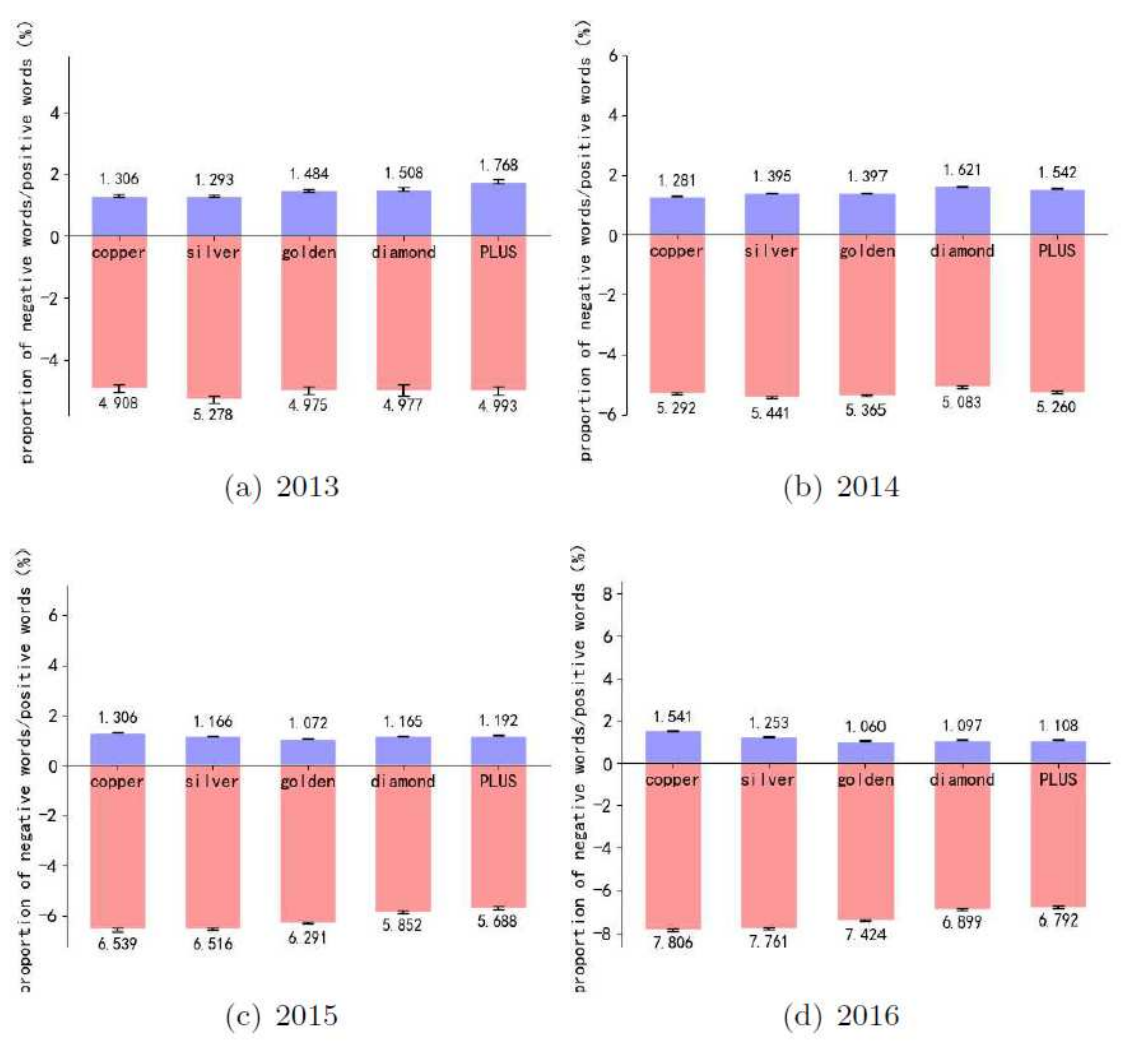}
\centering
\caption{\textbf{Average proportions for positive and negative emotion words in negative reviews of 2013-2016.}
      The bars on the upper side of the x-axis represent the percentage of positive emotion words, while those on the underside are the negative ones. The five groups, from left to right, refer to user levels, that is, copper, silver, golden, diamond and PLUS. The absolute value of the y-axis means the percentage of emotional words. Moreover, the error bar is calculated by the sample standard deviation. As the four bar plots show, the same tendency is performed in the negative degree and is regular in the positive degree in 2015-2017. Regarding the compared performances in consecutive years, it reflects that JD has emphasized the paid users' experience more and provided more rights or guarantees for them in recent years.}
\end{figure}

\newpage
\begin{figure}[H]
\scriptsize
\centering
\includegraphics[width=10cm]{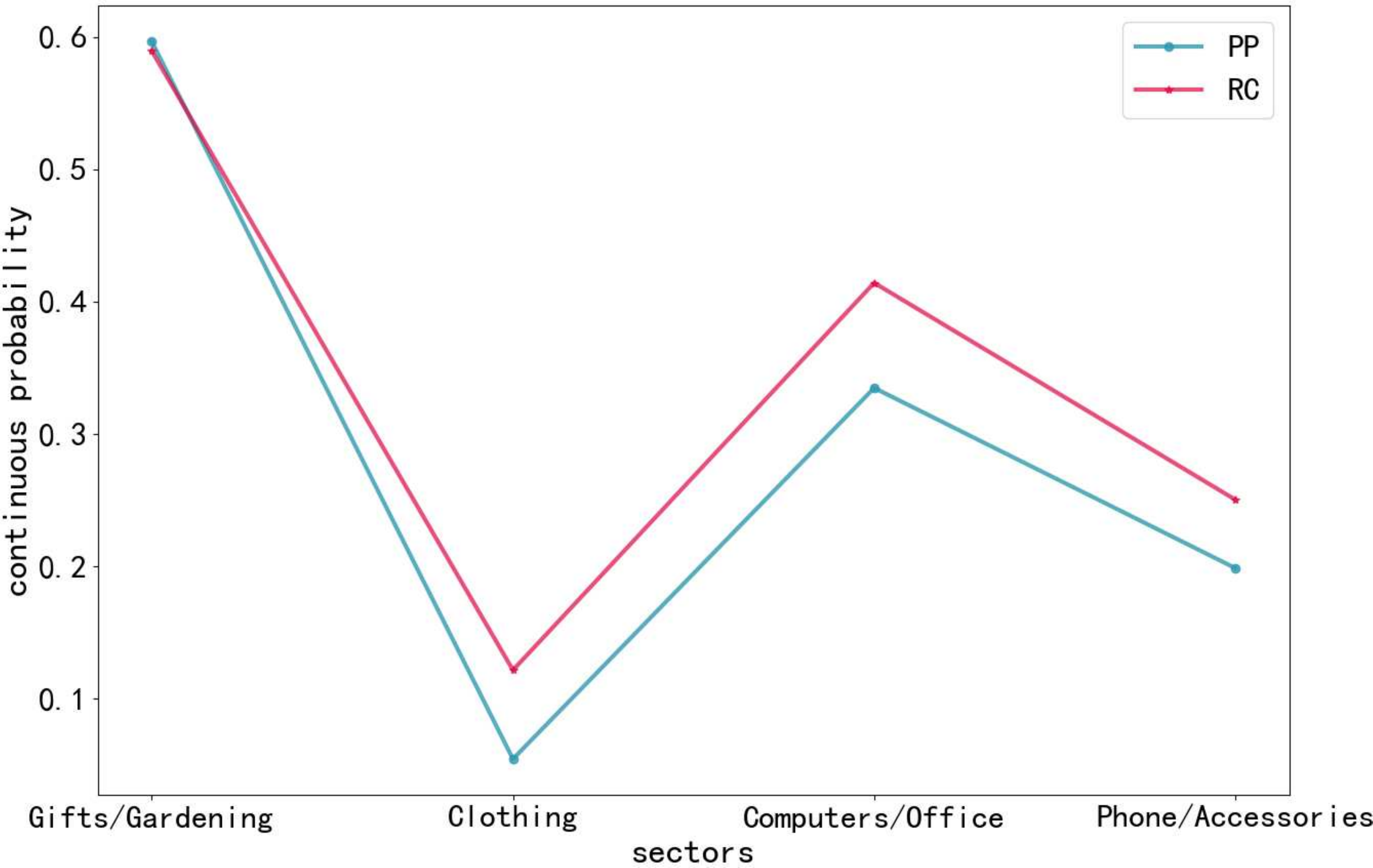}
\caption{\textbf{Continuous probability for PP time and RC time in a day of identified buyers.}
      The x-axis represents four sectors in our dataset, and the y-axis is the continuous probability. Moreover, it can be seen that the continuous probability for RC time is often larger than that for PP time.}
\end{figure}

\newpage
\begin{figure}[H]
\centering
\includegraphics[width=12cm]{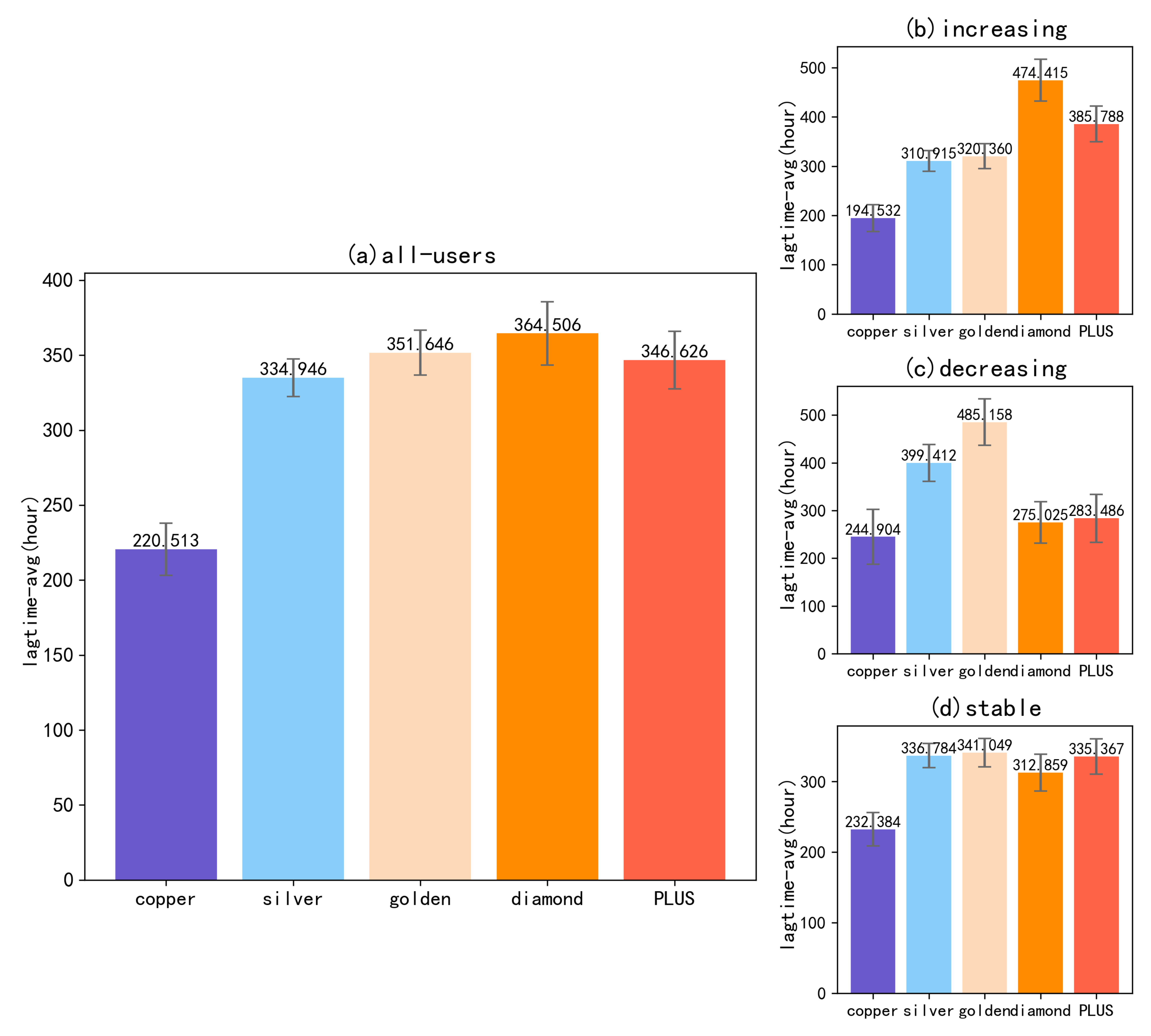}
\caption{\textbf{Average lags in an hour of five user levels' emotion shifts in all negative reviews, taking the Computer sector as an example.} The x-axis shows five user levels' emotion shifts. The height of different colors bars refers to the average lag of a certain kind of user levels' emotion shift in all after-use comments with initial scores equal to 1. Note that the results from other sectors are the same. Note that the upward trend of lags for increasing emotion in user level is statistically significant in the one-way $t$-test, which proves negative momentum again.}
\end{figure}

\newpage
\subsection*{Table A2: Topic words for three LDA models}
\label{Table A2}
\paragraph{Logistics; topic number=8}
\subparagraph*{${\rm topic}_1$}
'0.102*'express delivery' + 0.053*'poor' + 0.049*'delivery' + 0.042*'attitude' + 0.027*'service' 
+ 0.026*'staff' + 0.024*'bad' + 0.016*'slow' + 0.015*'service attitude' + 0.015*'thing' 
\subparagraph*{${\rm topic}_2$}
'0.054*'not received' + 0.032*'arrival' + 0.026*'goods' + 0.023*'thing' + 0.022*'signing' + 0.020*'receive' + 0.016*'fresh' + 0.016*'receipt' + 0.016*'display' + 0.012*'confirm'
\subparagraph*{${\rm topic}_3$}
'0.077*'package' + 0.026*'good' + 0.022*'poor' + 0.021*'express delivery' + 0.014*'opened' + 0.013*'carton' + 0.013*'box' + 0.010*'slowly' + 0.010*'wife' + 0.009*'buy'
\subparagraph*{${\rm topic}_4$}
'0.083*'bad review' + 0.017*'evaluation' + 0.016*'score' + 0.016*'express delivery' + 0.015*'bad' + 0.015*'one-score' + 0.015*'word' + 0.013*'do not want' + 0.010*'post' + 0.010*'sent'
\subparagraph*{${\rm topic}_5$}
'0.082*'express delivery' + 0.030*'deliver' + 0.029*'trash' + 0.024*'sent' + 0.023*'thing' + 0.020*'flowers' + 0.019*'call' + 0.018*'phone' + 0.016*'staff' + 0.014*'buy'
\subparagraph*{${\rm topic}_6$}
'0.192*'logistics' + 0.101*'too slow' + 0.070*'slow' + 0.028*'bad review' + 0.028*'speed' + 0.025*'ship' 
+ 0.016*'thing' + 0.013*'real thing' + 0.012*'Qixi Festival' + 0.010*'
(a title for one day in a week, such as Sunday/Monday...)'
\subparagraph*{${\rm topic}_7$}
'0.060*'arrival' + 0.025*'delivery' + 0.025*'deliver' + 0.020*'time' + 0.018*'deliver' + 0.018*'o'clock' + 0.018*'customer service' + 0.014*'night' + 0.014*'flowers' + 0.012*'next day'
\subparagraph*{${\rm topic}_8$}
'0.047*'date' + 0.044*'ship' + 0.032*'received' + 0.030*'buy' + 0.024*'order' + 0.022*'customer service' + 0.021*'goods' + 0.013*'seller' + 0.012*'flower' + 0.011*'thing'

\paragraph{Customer-service; topic number=9} 
\subparagraph*{${\rm topic}_1$}
'0.078*'bad review' + 0.026*'greeting card' + 0.022*'freight' + 0.016*'flower' + 0.015*'send' + 0.013*'broken' + 0.012*'corporation' + 0.010*'reimburse' + 0.009*'back' + 0.009*'screw'
\subparagraph*{${\rm topic}_2$}
'0.039*'customer service' + 0.014*'computer' + 0.014*'after-sales service' + 0.012*'buy' + 0.010*'girlfriend' + 0.010*'card' + 0.009*'phone' + 0.007*'ask' + 0.007*'hard disk' + 0.007*'look for'
\subparagraph*{${\rm topic}_3$}
'0.055*'return goods' + 0.035*'return' + 0.025*'seller' + 0.024*'apply for' + 0.021*'refund' + 0.017*'seller' + 0.017*'buy' + 0.016*'money' + 0.013*'reason' + 0.012*'received'
\subparagraph*{${\rm topic}_4$}
'0.219*'bad review' + 0.147*'invoice' + 0.028*'issue' + 0.018*'Valentine's Day' + 0.018*'flower' + 0.015*'billing' + 0.012*'received' + 0.011*'have been agreed' + 0.010*'customer service' + 0.009*'post'
\subparagraph*{${\rm topic}_5$}
'0.101*'customer service' + 0.027*'ask' + 0.022*'buy' + 0.016*'look for' + 0.015*'sent' + 0.013*'thing' + 0.013*'give away' + 0.012*'extremely bad' + 0.011*'received' + 0.010*'date'
\subparagraph*{${\rm topic}_6$}
'0.059*'bad' + 0.056*'service' + 0.041*'buy' + 0.037*'attitude' + 0.035*'thing' + 0.034*'customer service' + 0.030*'service staff' + 0.024*'trash' + 0.023*'poor' + 0.016*'after-sales service'
\subparagraph*{${\rm topic}_7$}
'0.023*'flower' + 0.016*'order(n)' + 0.013*'deliver' + 0.013*'Qixi Festival' + 0.012*'customer service' + 0.012*'seller' + 0.011*'delivery' + 0.010*'order(v)' + 0.009*'in advance' + 0.009*'order(v)'
\subparagraph*{${\rm topic}_8$}
'0.026*'flower' + 0.025*'shopping' + 0.022*'satisfactory' + 0.022*'customer service' + 0.019*'package' + 0.016*'fresh' + 0.011*'experience' + 0.009*'buy' + 0.009*'box' + 0.008*'purchase'
\subparagraph*{${\rm topic}_9$}
'0.032*'after-sales service' + 0.022*'computer' + 0.021*'buy' + 0.020*'install' + 0.017*'phone' + 0.016*'flower' + 0.015*'look for' + 0.012*'system' + 0.009*'not arrived' + 0.008*'paper'
\paragraph{Marketing; topic number=7} 
\subparagraph*{${\rm topic}_1$}
'0.075*'on sale' + 0.055*'buy' + 0.025*'just bought' + 0.023*'price' + 0.023*'computer' + 0.023*'price difference' + 0.022*'bad review' + 0.022*'finish' + 0.020*'the next day' + 0.015*'drop'
\subparagraph*{${\rm topic}_2$}
'0.036*'flower' + 0.025*'price' + 0.025*'yuan' + 0.023*'cheater' + 0.022*'card' + 0.020*'bad' + 0.018*'bad review' + 0.016*'buy' + 0.012*'expensive' + 0.011*'rose'
\subparagraph*{${\rm topic}_3$}
'0.112*'give away' + 0.031*'have been agreed' + 0.028*'gift' + 0.026*'flower' + 0.023*'buy' + 0.020*'bad review' + 0.014*'flower' + 0.011*'thing' + 0.010*'trash' + 0.010*'received'
\subparagraph*{${\rm topic}_4$}
'0.088*'buy' + 0.024*'flower' + 0.022*'thing' + 0.020*'defraud' + 0.019*'price' + 0.016*'on sale' + 0.013*'hold' + 0.011*'deceive' + 0.009*'real' + 0.007*'think'
\subparagraph*{${\rm topic}_5$}
'0.036*'customer service' + 0.014*'seller' + 0.012*'order(v.)' + 0.012*'fresh' + 0.012*'order(n.)' + 0.011*'date' + 0.011*'invoice' + 0.010*'flower' + 0.009*'buy' + 0.008*'ask'
\subparagraph*{${\rm topic}_6$}
'0.048*'real thing' + 0.036*'do not match' + 0.029*'consumer' + 0.021*'cheat' + 0.019*'commodity' + 0.018*'package' + 0.018*'description' + 0.016*'girlfriend' + 0.013*'photo' + 0.011*'color'
\subparagraph*{${\rm topic}_7$}
'0.028*'post' + 0.026*'seller' + 0.020*'buy' + 0.017*'give away' + 0.017*'greeting card' + 0.012*'fake' + 0.012*'Valentine's Day' + 0.011*'store owner' + 0.011*'cheat' + 0.009*'thing'\\
\footnotesize{$^a$ \textbf{For every topic, we output ten words with the top 10 largest weights in the model.} Given that the initial text is in Chinese, here, we translate the words into English. Considering that there are situations such as different Chinese words with the same or a close meaning, and the same Chinese word having different meanings in different situations, we sometimes translate them into the same word or indicate the property of the word in the bracket after it, such as $n,v$.}\\
\end{document}